\documentclass[aps,pre,graphicx,reprint,amssymb,showpacs,superscriptaddress,floatfix,nobalancelastpage]{revtex4-1}
\usepackage[utf8]{inputenc}
\usepackage[T1]{fontenc}
\usepackage[english]{babel}
\usepackage{amsmath}
\usepackage{amsfonts}
\usepackage{amssymb,textcomp}
\usepackage[xcdraw]{xcolor}
\usepackage{colortbl}
\usepackage{tikz}
\usepackage{bm}
\usepackage{framed}
\usepackage{graphicx, wrapfig}
\usepackage{array,multirow}
\usepackage{lmodern}					
\usepackage{comment}
\usepackage{natbib}
\usepackage[margin=2cm]{geometry}
\usepackage{hyperref}

\usepackage{scalerel}
\usepackage{enumitem}

\newcommand\beq{\begin{equation}}      
\newcommand\beqnn{\begin{eqnarray*}}   
\newcommand\beqa{\begin{eqnarray}}     
\newcommand\beqann{\begin{eqnarray*}}  

\newcommand\eeq{\end{equation}}        
\newcommand\eeqnn{\end{eqnarray*}}     
\newcommand\eeqa{\end{eqnarray}}       
\newcommand\eeqann{\end{eqnarray*}}    

\newcommand{\ave}[1]{\left\langle #1 \right\rangle}

\newcommand{\modsq}[1]{\left| #1 \right|^{2}}                                               
\def\nl {\nonumber \\}
\def\nln {\\}

\newcommand\bi{\begin{itemize}}
\newcommand\ei{\end{itemize}}

\def\nl {\nonumber \\}

\def\bx{\bm x}

\newcommand{\eref}[1]{(\ref{#1})}

\newcommand{\Eref}[1]{Eq.~(\ref{#1})}

\newcommand{\al}[1]{\begin{align} #1 \end{align}}
\def\l0{L}

\def\xt{\bm x,t}
\def\lama{\lambda} 
\def\modx{|\bm x|}
\def\lamt{\tilde\lambda} 
\def\xit{\tilde\xi} 
\def\L0{\bm L_0} 
\def\Lt{\bm L(t)} 
\def\lt{\ell_t}

\renewcommand{\vec}[1]{\bm{#1}}

\newcommand*\colvec[3][]{
    \begin{pmatrix}\ifx\relax#1\relax\else#1\\\fi#2\\#3\end{pmatrix}
}
\makeatletter
\def\oversortoftilde#1{\mathop{\vbox{\m@th\ialign{##\crcr\noalign{\kern3\p@}%
      \sortoftildefill\crcr\noalign{\kern3\p@\nointerlineskip}%
      $\hfil\displaystyle{#1}\hfil$\crcr}}}\limits}

\def\sortoftildefill{$\m@th \setbox\z@\hbox{$\braceld$}%
  \braceld\leaders\vrule \@height\ht\z@ \@depth\z@\hfill\braceru$}

\makeatother

\begin{document}

\title{Correlations and forces in sheared fluids with or without quenching}
\author{Christian M. Rohwer}
\email[]{crohwer@is.mpg.de}

\affiliation{Max Planck Institute for Intelligent Systems, Heisenbergstr. 3, 70569 Stuttgart, Germany}
\affiliation{4th Institute for Theoretical Physics, University of Stuttgart, Pfaffenwaldring 57, 70569 Stuttgart, Germany}

\author{Anna Maciolek}
\affiliation{Max Planck Institute for Intelligent Systems, Heisenbergstr. 3, 70569 Stuttgart, Germany}
\affiliation{4th Institute for Theoretical Physics, University of Stuttgart, Pfaffenwaldring 57, 70569 Stuttgart, Germany}
\affiliation{Institute of Physical Chemistry, Polish Academy of Sciences, Kasprzaka 44/52, PL-01-224 Warsaw, Poland}

\author{S. Dietrich}
\affiliation{Max Planck Institute for Intelligent Systems, Heisenbergstr. 3, 70569 Stuttgart, Germany}
\affiliation{4th Institute for Theoretical Physics, University of Stuttgart, Pfaffenwaldring 57, 70569 Stuttgart, Germany}
 
\author{Matthias Kr\"uger}
\affiliation{Institute for Theoretical Physics, University of G\"ottingen, D-37077, G\"ottingen, Germany}

\date{\today}

\begin{abstract}
Spatial correlations play an important role in characterizing material properties related to non-local effects. Inter alia, they can give rise to fluctuation-induced forces. Equilibrium correlations in fluids provide an extensively studied paradigmatic case, in which their range is typically bounded by the correlation length. Out of equilibrium, conservation laws have been found to extend correlations beyond this length, leading, instead, to algebraic decays. In this context, here we present a systematic study of the correlations and forces in fluids driven out of equilibrium simultaneously by quenching and shearing, both for non-conserved as well as for conserved Langevin-type dynamics. We identify which aspects of the correlations are due to shear, due to quenching, and  due to simultaneously applying both, and how these properties depend on the correlation length of the system and its compressibility. Both shearing and quenching lead to long-ranged correlations, which, however, differ in their nature as well as in their prefactors, and which are mixed up by applying both perturbations.
These correlations are employed to compute non-equilibrium fluctuation-induced forces in the presence of shear, with or without quenching, thereby generalizing the framework set out by Dean and  Gopinathan. 
These forces can be stronger or weaker compared to their counterparts in unsheared systems. In general, they do not point along the axis connecting the centers of the small inclusions considered to be embedded in the fluctuating medium. 
Since quenches or shearing appear to be realizable in a variety of systems with conserved particle number, including active matter, we expect these findings to be relevant for experimental investigations. 
\end{abstract}

\maketitle

\section{Introduction}
\label{sec:intro}

Long-ranged correlations (LRCs) play an important role in both the static and dynamic properties of many-body systems~\cite{onukibook,kardarbook,taeuberbook}. For example, they can generate  so-called fluctuation-induced forces~\cite{kardargolestanian1999}. The latter have been studied and observed in the setting of electromagnetic fields \cite{casimir1948,bordag} or in classical systems \cite{fisherdegennes,krechbook,hertlein2008,gambassiCCFPRE2009,garciachan2002,garciachan2006,fukutowettingfilms2005,linzandi2011}. A prominent example, in which LRCs occur, is a system near a second-order phase transition. In anisotropic systems, asymmetric objects may also experience Casimir torques \cite{kondrat2009critical}.

In out-of-equilibrium systems LRCs are more common~\cite{grinsteinleesachdev1990}; they are typically related to conservation laws (e.g., conserved particle number or momentum), as demonstrated in various systems \cite{spohn1983,dorfmankirkpatricksengers1994,mukamelkafri1998,dorfmankirkpatricksengers1994,croccolo2016non,oshanindemery2017prl}. These non-equilibrium LRCs, in turn, give rise to associated non-equilibrium fluctuation-induced forces. Such forces have been studied theoretically for systems with gradients in temperature \cite{kirkpatricksengers2013,kirkpatrick2015prl,kirkpatrick2016pre} or density \cite{aminovkardarkafri2015}, quenched systems \cite{rohwer2017transient, rohwer2018forces}, stochastically driven systems~\cite{Mohammadi17}, in systems under shear \cite{Gompper2017shear, KirkpatrickSengers2018shear}, and within fluctuating hydrodynamics \cite{Monahan16}. 

Here, we aim at studying correlations and forces in fluid systems undergoing up to two non-equilibrium perturbations simultaneously, i.e., shearing and quenching. In pursuit of correlations which extend far beyond microscopic length scales, we resort to the well-known coarse-grained dynamical models: ``model A'' (describing a non-conserved field) and ``model B'' (describing a conserved field) \cite{hohenberg,kardarbook}. These models have  been applied extensively in describing various dynamical situations, e.g., the approach of the critical point from non-equilibrium initial conditions~\cite{gambassi2008EPJB,deangopinathan2009JStatMech,deangopinathan2010pre,Gross2018Surface}, the coarsening following a temperature quench~\cite{sutapa1,sutapa2}, or for driven systems at criticality \cite{demery2010,gambassi2013prl}. They have also been used to study shearing of near-critical fluids \cite{onukikawasaki_fluctuations_1978,onukikawasaki_NESS-RG_1979, corberi1998, corberi1999,Gonella2000,corberi2003,rohwer2017viscosity}, leading to a large variety of phenomena.

The use of such models provides generic scenarios, which we expect to be relevant for physical systems which allow for shearing and/or quenching. Shear is directly experimentally accessible  \cite{larson}. Quenches can also be realized, for instance by using effective interactions of particles which can be changed suddenly, e.g., by swelling particles \cite{ballauff2006} or through external fields \cite{maretkeim2004}. Another type of quench concerns a sudden change of temperature, which is a perturbation often employed in order to obtain supercooled liquids \cite{Debenedetti01}. Such quenches of temperature (or of noise strength) may be achieved experimentally also in active fluids \cite{solon_active_2015, berthier2015epl,catesX}, which in many respects can be described by the use of effective temperatures \cite{loiEffectiveT2008, bizonne, rohwer2018forces}.

The manuscript is structured as follows. In Sec.~\ref{sec:sys} we begin with a detailed description of the system as well as of the model under consideration. Post-quench correlations in the absence of shear are briefly reviewed in Sec.~\ref{sec:review}. In Sec.~\ref{sec:SS}, this is followed by an analysis of the effect of conservation laws on steady state correlations of weakly sheared systems. The case of dissipative dynamics is discussed in Sec.~\ref{sec:modA}, while Sec.~\ref{sec:modBSS} deals with conserved density fluctuations. The dependence of the  (equal-time) correlation function on space and time is computed for model B in Sec.~\ref{sec:modBDyn}. Using the formal solution derived in Sec.~\ref{sec:formalsol}, this quantity can be determined analytically in certain limits (Sec.~\ref{sec:modBDynLocal}). Correlations between points advected by the shear field are discussed in Secs.~\ref{sec:modBDynComov} and \ref{sec:modBDynNonLoc} for various limiting cases. Section~\ref{sec:incsec} presents a formalism for computing non-equilibrium fluctuation-induced forces in the presence of quenching and shearing. This extends the framework of Ref.~\cite{deangopinathan2010pre} to include shear. While this formalism holds for various geometries which do not couple to the shear flow, such as films formed by parallel plates, it is employed in Sec.~\ref{sec:FSS} in order to compute forces between small inclusions embedded in steadily sheared systems, as well as for dynamic post-quench forces (PQFs) under shear (Sec.~\ref{sec:Faq}). In Table \ref{tab_glossary} we provide a glossary of commonly used quantities.

\section{Physical system and model}
\label{sec:sys}

\begin{table*}[t]
\begin{center}
\begin{tabular}{c c c}
\hline\hline
Quantity & Description & Definition in \\
\hline
$\mu_{A/B}$ & mobility coefficient for model A/B  &  Sec.~\ref{sec:cg}\\ 
$\xi$ & inherent (equilibrium) correlation length of the fluid&  Sec.~\ref{sec:cg}\\ 
$m$ & ``mass'' / compressibility coefficient &  Sec.~\ref{sec:cg}\\ 
$D$ & diffusion coefficient: $D = \mu_A m \xi^2$ (model A) or $D = \mu_B m$ (model B)&  \Eref{eq:Ddef}\\ 
$\dot\gamma$ & shear rate of imposed flow with $\bm v = \dot\gamma y\bm e_x$  &  Sec.~\ref{sec:sysdyn}\\ 
$T_I,\, T$ & temperatures before ($t<0$) and after ($t\geq0$) the quench, respectively &  Sec.~\ref{sec:review}\\ 
$\lt=\sqrt{Dt}$ & diffusion-induced length scale&  Sec.~\ref{sec:lengthscales}\\ 
$\lambda=\sqrt{D/\dot\gamma}$ & shear-induced length scale&  Sec.~\ref{sec:lengthscales}\\ 
$\phi(\xt)
$ & fluctuations of the density $\rho(\xt)$ about its mean value $\ave{\rho(\xt)}$&  Sec.~\ref{sec:cg}\\ 
$C(\xt) 
$ & equal-time correlation function of $\phi(\xt)$ 
in the bulk&  \Eref{eq:Cxtgen}\\ 
$\modx$ & distance between points in bulk; observation length scale for correlations &  \Eref{eq:Cxtgen}\\ 
$\bm x_0$, $\bm x(t)$ & vector between two fixed or two co-moving points in the shear flow, respectively &  Sec.~\ref{sec:modBDynComov}\\ 
$a^{(\bm u)}$ & for any vector $\bm u = (u_x,\bm u^\perp)$, $a \equiv|\bm u^\perp|/|\bm x|$
& Sec.~\ref{sec:modBDynComov} \\ 
$\Omega^{(\bm x)}_{\alpha} = (\hat{\bm x})_\alpha
$ & $\alpha$-component of the unit vector $\hat{\bm x} = \vec{x}/|\bm x|$ & \Eref{eq:Omegadef}\\ 
$t_\ell^* = D t / \ell^2$ & dimensionless diffusive time across the distance $\ell\in\{\modx,|\L0|\}$&  
\Eref{eq:tstardef}\\ 
$\bm L$ & vector connecting two stationary ($\bm L=\L0$) or co-moving ($\bm L=\bm L(t)$) inclusions in shear flow  &  Sec.~\ref{sec:incsec}\\
$\lamt,\xit$ & $\lambda,\xi$ rescaled by $\modx$ (correlation functions) or $|\L0|$ (forces) &  \Eref{eq:scalvars}, Sec.~\ref{sec:incsec}\\ 
$\bm F_{\textrm{A/B}}(t\!\to\!\infty)$ & steady-state force (in model A/B) between two inclusions in shear flow  &  Sec.~\ref{sec:FSS}\\
$\bm F_{\textrm{s}}(\bm L_0,t)$ & post-quench force (PQF) between two stationary inclusions separated by $\L0$  &  Sec.~\ref{sec:PQFfixed}\\
$\bm F_{\textrm{c-m},\dot\gamma}(\bm L(t),t)$ & PQF between two co-moving  inclusions in a sheared fluid 
&  Sec.~\ref{sec:Fcomovsec}\\
$\bm F_{\textrm{c-m},0}(\bm L(t),t)$ &  PQF between inclusions following the co-moving trajectory $\bm L(t)$, but fluid is unsheared
&  Sec.~\ref{sec:Fcomovsec}\\
\hline\hline
\end{tabular}
\end{center}
\caption{Glossary of the quantities most frequently used in the present study. }
\label{tab_glossary}
\end{table*}

\subsection{Coarse-grained model: equilibrium properties}
\label{sec:cg}
Aiming at the analysis of correlations in classical fluids, which extend far beyond microscopic length scales, we employ classical field theory based on the Landau-Ginzburg theory for a scalar order parameter field $\phi$ \cite{onukibook,kardarbook}. With a one-component fluid in mind, $\phi$ describes density fluctuations $\phi(\xt) = \rho(\xt) - \ave{\rho(\xt)}$,  where $\rho(\xt)$ is the (snapshot) number density distribution. In the case of binary liquid mixtures (in the mixed state), $\rho(\xt)$ corresponds to a local concentration of the particles. The vector $\vec{x}$ is a $d$-dimensional position vector, and $t$ denotes time. 
Thermodynamically far from phase transitions, a  Gaussian Hamiltonian $H$ is expected to capture the leading influence of the fluctuations. Thus we consider
\al{
H=\int \textrm d^dx \left[\frac \kappa 2 (\nabla\phi)^2+\frac m 2 \phi^2(\bm x)\right],
\label{eq:H}
}
which induces a (bulk) correlation length $\xi = \sqrt{\kappa/m}$. In thermal equilibrium and for $d>2$, the Hamiltonian in Eq.~\eqref{eq:H} gives rise to the following two-point correlation function~\cite{kardarbook}:
\al{
\ave{\phi(\bm x)\phi(\bm 0)}^{\textrm{eq}} = \frac{k_B T}m
\begin{cases}
\frac{\modx^{2-d}}{(2-d)S_d\xi^2}, &|\bm x|\ll\xi\\
\frac{\xi^{-(d+1)/2}e^{-\modx/\xi}}{(2-d)S_d \modx^{(d-1)/2}}, &|\bm x |\gg\xi,
\end{cases}
\label{eq:Ceq}
}
where $S_d = {2\pi^{(d+1)/2}}/{\Gamma(\frac{d+1}{2})}$ is the surface area of a $d$-dimensional unit sphere. $T$ is the temperature and $k_B$ is the Boltzmann constant. For systems far away from phase transitions, $\xi$ is small so that the lower line in \Eref{eq:Ceq}, i.e.,  $|\bm x |\gg\xi$ applies. In this regime, the correlation function decays exponentially as a function of $x/\xi$. (Here we do not consider the presence of long-ranged forces such as van der Waals forces, which asymptotically give rise to an algebraic decay even for $|\bm x |\gg\xi$.) Within the Gaussian approximation, $m$ in Eq.~\eqref{eq:H} can be expressed in terms of the isothermal compressibility $\chi_T = - (\partial V / \partial P)_T/V$ \cite{Chandler93,krugerdean2017a, HansenMcDonald} ($V$ is the system volume):  
\begin{equation}
  \label{eq:mass}
  m=\frac{1}{\rho_0^2\chi_T},
\end{equation}
where $\rho_0$ is the mean bulk density.

\subsection{Dynamical description with shear}
\label{sec:sysdyn}
\subsubsection{Equations of motion}
The dynamical description employed here is based on the Hamiltonian given in Eq.~\eqref{eq:H} and on Langevin equations for the field $\phi$ within model A and model B \cite{hohenberg}. These models consider non-conserved and conserved dynamics, respectively. As far as shear is concerned, we consider a simple shear velocity profile $\bm v$, so that any feedback effects of the field $\phi$ onto the velocity profile, as well as fluctuations of $\bm v$, are neglected (in contrast to  model H \cite{onukibook}, which includes these couplings).
Using $\bm v = \dot\gamma y \bm e_x$ with shear rate $\dot\gamma \equiv \partial v_x / \partial y$~\cite{larson}, the Langevin equation reads
\al{
\partial_t \phi  +\dot \gamma y \frac{\partial \phi}{\partial x}&= \hat\mu(\kappa \nabla^2 -m) \phi + \eta(\bm x, t),
\label{eq:Langevin}
}
where the white noise obeys the spatio-temporal correlations
\al{
\ave{\eta(\bm x, t)\eta(\bm x', t') } &= 2\hat\mu k_B T \delta(\bm x - \bm x') \delta(t-t').
\label{eq:noisecorr}
}
The mobility operator $\hat \mu$ encodes whether $\phi$ is conserved or not:
\al{
\hat\mu =
\begin{cases}
\mu_A,\quad &\textrm{model A},\\
-\mu_B\nabla^2,\quad &\textrm{model B}.
 \end{cases}
\label{eq:hatGamma}
}
We note that the coefficients $\mu_{A/B}$ carry different dimensions. 
Defining the Fourier transform $\mathcal F$ as $f(\bm k)=\mathcal F[f](\bm k)= (2\pi)^{-d}\int \textrm d ^d k \exp(i \bm k \cdot \bx) f(\bx)$, Eqs.~\eref{eq:Langevin} and \eref{eq:noisecorr} can be expressed in Fourier space:
\al{
\partial_t \phi(\bm k, t)&= \hat O(\bm k,\{\partial_{k_i}\}) \phi(\bm k, t) + \eta(\bm k, t), \nl
\ave{\eta(\bm k, t)\eta(\bm k', t') } &= 2(2\pi)^{d}\mu_k k_B T \delta(\bm k+ \bm k') \delta(t-t'),
\label{eq:LangevinFourier}
}
where we have introduced the operator
\al{
\hat O(\bm k,\{\partial_{k_i}\}) \equiv \dot \gamma k_x \partial_{k_y} -\mu_k(\kappa k^2 + m)
\label{eq:Odef}
} 
and represented $\hat \mu(\bm x)$ from \Eref{eq:hatGamma} in terms of its spectrum
\al{
\mu_k =
\begin{cases}
 \mu_A,\quad &\textrm{model A},\\
 \mu_B k^2,\quad &\textrm{model B}.
\end{cases}
}
The quantity of our interest is the time-dependent structure factor $C(\bm k, t)$. It is defined as
\al{
\ave{\phi(\bm k, t)\phi(\bm k', t)}
&\equiv (2\pi)^d\delta^d(\bm k + \bm k') C(\bm k, t).
\label{eq:Cktdef}
}
$C(\bm k, t)$ depends on time because the system is out of equilibrium. It is evaluated at equal times, and evolves according to
\al{
\partial_t C(\bm k,t)  =2\hat O(\bm k,\{\partial_{k_i}\})C(\bm k,t)+ 2  k_B T \mu_k,
\label{eq:dtCk}
}
which has the general solution
\al{
C(\bm k, t) = &e^{2t \hat O(\bm k,\{\partial_{k_i}\})}C(\bm k, t=0) \nl
&\quad+ 2 k_B T \int_0^{t} ds\; e^{2(t-s)\hat O(\bm k,\{\partial_{k_i}\})}\mu_k,
\label{eq:Cktsol}
}
where $C(\bm k,t= 0)$ is the structure factor at $t=0$. Importantly, $\hat O(\bm k,\{\partial_{k_i}\})$ comprises powers of $k_i$  and $\partial_{k_i}$ with $i = 1\dots d$, and therefore the exponents must be expanded using the Zassenhaus formula \cite{casasZassenhaus2012}. Expressions such as in \Eref{eq:Cktsol} have been discussed in the literature --- see, e.g., Refs.~\cite{onukikawasaki_fluctuations_1978,onukikawasaki_NESS-RG_1979,corberi1998,corberi1999,Gonella2000,corberi2003}. (We note that there are discrepancies of a factor of 2 among the latter references regarding the coefficients of the terms in \Eref{eq:dtCk}. According to our derivation, \Eref{eq:dtCk}, which follows directly from the Langevin \Eref{eq:LangevinFourier}, fixes these constants via \Eref{eq:Odef}.) However, our aim is to obtain explicit expressions \textit{in position space}: using Eqs.~\eref{eq:Cktdef} and \eref{eq:Cktsol}, the time-dependent equal-time correlation function 
\al{
C(\bm x ,t) \equiv \ave{\phi(\bm x,t)\phi(0,t)}
\label{eq:Cxtgen}
}
can be found by Fourier inversion. Here $\modx$ is the distance in the bulk between two points the correlation of which is being considered.

\subsubsection{Quenching at $t=0$}
We consider the dynamics defined by Eqs.~\eref{eq:Langevin} and \eref{eq:noisecorr} subject to a quench at time $t=0$; this amounts to a sudden change of one or more of the parameters in these equations. For instance, this parameter can be the temperature $T$. Such a description in terms of instantaneous changes of parameters is based on the assumption that processes at small length scales relax on short time scales, so that the mesoscopic parameters rapidly attain their new values. Thus, in the general solution given by Eq.~\eqref{eq:Cxtgen}, the first term on the rhs depends on the parameter values before the quench (in the following denoted with subscript $I$), while the second term on the rhs depends on the parameter values after the quench (for which no subscript is used).

Physically, the parameters in Eqs.~\eref{eq:Langevin} and \eref{eq:noisecorr} are in general not independent; for example, a change in temperature may also change the coefficent $m$ via Eq.~\eqref{eq:mass}. However, we treat these quantities as being independent, thereby allowing for a wide variety of quenching scenarios. 

Regarding Eq.~\eqref{eq:Ceq}, we note that, in the absence of shear, the correlation function depends on the ratio $k_BT/m$ and on the correlation length $\xi$. Therefore a quench induces a non-equilibrium, transient dynamics if one of these parameters is changed.  

\subsubsection{Important length scales}
\label{sec:lengthscales}
It is useful to introduce the collective diffusion coefficient $D$ \cite{dhont}, which follows from \Eref{eq:Langevin}:
\al{
D =
\begin{cases}
 \mu_A\kappa = \mu_A m\, \xi^2,\quad &\textrm{model A},\\
 \mu_B m,\quad &\textrm{model B}.
\end{cases}
\label{eq:Ddef}
}
We recall that, in model A, $D$ vanishes in the limit $\xi\to0$, because in the absence of correlations, the relaxation mechanism of model A is local and not diffusive.  $D$ gives rise to two length scales:
\al{
\lambda= \sqrt{{D}/{\dot\gamma}} \quad \textrm{and} \quad \lt = \sqrt{D t}.
\label{eq:lambdadef}
}
Here $\lambda$ is the length scale on which shear and diffusion have comparable effects, i.e., regions with $|\vec{x}|\ll \lambda$ are diffusion-dominated, while regions with $|\vec{x}|\gg \lambda$ are shear-dominated. The quantity $\lt$ is the typical distance covered by diffusion within the time $t$. 

Thus the Langevin equation \eref{eq:Langevin} depends on the length scales $|\bm x|$, $\lt$, $\lambda$, and $\xi$, where $\modx$ is the observation scale of a given observable. Regarding notation, we shall employ vectors $\bm x$ when referring to points in the bulk, and vectors $\bm L$ when denoting vectors connecting external objects immersed in the fluid (e.g., for computing forces between certain objects in Sec.~\ref{sec:incsec}).

\section{Quench in the absence of shear}
\label{sec:review}
Here, we briefly recall the main findings of Refs.~\cite{rohwer2017transient, rohwer2018forces}, in which quenches in the absence of shear were studied.

The explicit evaluation of Eqs.~\eqref{eq:Cktsol} and \eqref{eq:Cxtgen} for $\dot\gamma=0$ yields, within model B and to leading order in $\xi$ (recall that indices $I$ denote parameters before the quench),
\begin{align}
\ave{\phi(\bx,t)\phi(0,t)}
&= \Big[\frac{k_B T_I}{m_I} - \frac{k_B T}{m}\Big] \frac{1 }{ \modx^d }\frac{e^{-\frac{1}{8 t^*}}}{(8 \pi  t^*)^{d/2}}.
\label{eq:CPRL}
\end{align}
The dimensionless quantity
\al{
t^*_{\modx} = D t / \modx^2= \lt^2/ \modx^2
\label{eq:tstardef}
}
is obtained by rescaling time by the diffusive time scale across the distance $|\bm x|$ [see \Eref{eq:lambdadef}]. Equation~\eqref{eq:CPRL} shows that a quench gives rise to non-equilibrium LRCs, which, by virtue of their algebraic spatial decay, extend beyond the correlation length $\xi$. These LRCs are, to leading order in $\xi$, independent of $\xi$. The rescaled time [compare Eqs.~\eqref{eq:CPRL} and \eqref{eq:Ceq}] illustrates that $\lt =\sqrt{Dt}$ plays the role of a time-dependent correlation length. For long times, the correlation function in Eq.~\eqref{eq:CPRL} decays algebraically in time, as the system approaches the new equilibrium state.

The result within model A is qualitatively different in that the range of the correlations is restricted by $\xi$, so that for $\modx\gg \xi$, de facto no correlations are present. Therefore the conservation law associated with model B is the key ingredient which explicitly gives rise to the aforementioned non-equilibrium LRCs.

\section{Perturbative analysis of weak shear in steady state}
\label{sec:SS}

Having reviewed the quenching process without shear in Sec.~\ref{sec:review}, we proceed by analyzing the case of shear without quenching, i.e., the case of a steadily sheared system.

\subsection{Non-conserved density fluctuations: model A}
\label{sec:modA}
Since solving Eq.~\eqref{eq:dtCk} for arbitrary $\dot\gamma$ is challenging, we treat shearing perturbatively, i.e., we expand the correlation function according to
\al{
C= C^{(0)}+\dot\gamma C^{(1)}+ \mathcal{O}(\dot\gamma^2),
\label{eq:Cexpand}
} 
with $C^{(0)}$ and $C^{(1)}$ evaluated at $\dot\gamma =0$, i.e., without shearing. Such an expansion is valid if the length scale $\lambda$ in Eq.~\eqref{eq:lambdadef} is the largest one to be considered, i.e., 
\al{
\lambda\gg \{|\bm x| , \lt, \xi\}.
\label{eq:linresplengths}
}
Thus, the above expansion in terms of powers of the shear rate is valid for small observation scales $\modx$, short times $t$, and small correlation lengths $\xi$. Since here we consider steady states (i.e., times long after the quench), $\lt$ in Eq.~\eqref{eq:linresplengths} is replaced by $\ell_\tau$, where $\tau$ is the time scale for the relaxation of density fluctuations in the system.

The structure factor $C(\bm k,t)$ obeys the differential equation \eref{eq:dtCk} with $\mu_k = \mu_A$, and the steady state can be obtained from the limit $t\!\to\!\infty$. $C^{(0)}$ is found as \cite{kardarbook}
\al{
C^{(0)}(\bm k,t\!\to\!\infty) 
=\frac{   k_B T}{\kappa k^2 +m} .
\label{eq:CktModAzeroshear}
}
Fourier inversion of Eq.~\eqref{eq:CktModAzeroshear} yields Eq.~\eqref{eq:Ceq} above.

The contribution linear in $\dot\gamma$ follows from re-inserting Eq.~\eqref{eq:CktModAzeroshear} into \Eref{eq:dtCk}, yielding
 $C^{(1)}= \frac{k_x}{\mu_A (\kappa k^2 +m)}\partial_{k_y}C^{(0)}$, i.e.,
\al{
\frac{\dot\gamma C^{(1)}(\bm k,t\!\to\!\infty)}{ k_B T} 
= -\frac{\dot\gamma \kappa }{\mu_A}  \frac{2 k_x k_y}{\left(\kappa k^2 +m\right)^3}.
\label{eq:CkmodASSk}
}
In $d=3$, this can be Fourier-inverted, yielding
\al{
\frac{ \dot\gamma C^{(1)}(\bm x,t\!\to\!\infty)}{ k_B T}
&=\frac{\dot \gamma }{16 \pi  \kappa ^2 \mu_A}\frac{x y \exp\left({-\left| \bm x\right| \sqrt{{m}/{\kappa }}}\right)}{\left| \bm x\right|} \nl
&= 
\frac{\Omega_x^{(\bm x)}\Omega_y^{(\bm x)}}{16 \pi  m} \frac{e^{-1/\tilde \xi}}{\tilde\lama^2 \tilde \xi^2}  \frac{1}{\left| \bm x\right|^3}.
\label{eq:CkmodASSx}
}
Here
\al{
\Omega_\alpha^{(\bm x)} = \bm x_\alpha/|\bm x|
\label{eq:Omegadef}
} 
is the $\alpha$-component of the unit vector $\hat{\bm x} = \vec{x}/\modx$. In $d=3$, for instance, $\Omega_x = \sin\vartheta\cos\varphi$, $\Omega_y = \sin\vartheta\sin\varphi$, and  $\Omega_z = \cos\vartheta$, in terms of the polar (azimuthal) angle $\varphi\in[0,2\pi]$ ($\vartheta\in[0,\pi]$). We have also introduced the rescaled lengths
\al{
	\tilde \xi= {\xi}/{\modx}\quad\textrm{and} \quad\tilde\lambda = {\lambda}/{\modx}.
\label{eq:scalvars}
}
(In Sec.~\ref{sec:incsec}, where we shall study external objects (initially) separated by a vector $\bm L_0$, $\Omega_\alpha$ refers to the angles of $\bm L_0$, and the quantities $\tilde \xi$ and $\lamt$ are understood to be scaled by $|\bm L_0|$ instead of $\modx$.)

Equation \eref{eq:CkmodASSx} illustrates that shear induces a correction to \Eref{eq:Ceq}, which, just as the equilibrium result, decays exponentially on the length scale $\xi$, so that shear amounts to a quantitative, but not qualitative correction. Note, however, that the algebraic prefactors of ${e^{-1/\tilde \xi}}$ for $C^{(0)}$ and $C^{(1)}$ are given by $\modx^{-1}$ and $\modx^{-3}$, respectively. Furthermore, this correction vanishes for $\xi\!\to\!\infty$. It is worth noting that for model A this limit does not contradict Eq.~\eqref{eq:linresplengths}, because $\lama = \xi \sqrt{\mu_A m/\dot\gamma}$ is proportional to $\xi$.

\subsection{Conserved density fluctuations: model B}
\label{sec:modBSS}
In the case of model B dynamics, $C(\bm k,t)$ obeys \Eref{eq:dtCk} with $\mu_k = k^2m\mu_B$. The expression for zero shear is identical to Eq.~\eqref{eq:CktModAzeroshear}, reflecting the fact that the choice of the dynamic model has no influence on the equilibrium (Boltzmann) distribution. The term linear in the shear rate follows from the perturbative expansion of \Eref{eq:Cexpand}, which in this case gives $C^{(1)} = \frac{k_x}{\mu_B k^2(\kappa k^2 +m)}\partial_{k_y}C^{(0)}$, i.e.,
\al{
\frac{\dot\gamma C^{(1)}(\bm k,t\!\to\!\infty)}{ k_B T} 
&=-\frac{\dot\gamma \kappa }{\mu_B}  \frac{2 k_x k_y}{k^2\left(\kappa k^2 +m\right)^3}. 
\label{eq:CkmodBSS}
}
As above, this expression can be Fourier inverted analytically for $d=3$ \footnote{
We make use of the Fourier sine transform for a spherically symmetric function $f(\bm k) = f(k)$,
$\int {\frac{1}{{\left( {2\pi } \right)^3 }}{e^{i\vec k \cdot \vec r} }f(\vec k) d^3 \vec k}  
= \frac{2}{{\left( {2\pi } \right)^2 }}\int\limits_0^\infty  {k^2 f(\vec k )\frac{{\sin \left( {kr} \right)}}{{kr}}dk}$
and the fact that $\mathcal F^{-1}[k_x k_y g(\bm k)](\bm x) = -\partial_x \partial_y \mathcal F^{-1}[ g(\bm k)](\bm x)$.
}, yielding
\al{
&\frac{\dot\gamma C^{(1)}(\bm x,t\!\to\!\infty)}{ k_B T}
=\frac{\Omega^{(\bm x)} _x \Omega^{(\bm x)} _y}{\pi m  \modx^3  } \frac{e^{-1/\xit}}{\lamt ^2} \nl 
&\qquad\times \Big[
\frac{3 \left(e^{1/\xit}-1\right)\xit}{2 } -\frac{3\xit}{2 } -\frac{3}{4} -\frac{1}{4\xit} -\frac{1}{16\xit^2}
\Big]. 
\label{eq:CkmodBSSlinshear}
}
We recall from \Eref{eq:Ceq} that equilibrium correlations decay exponentially for $\modx\gg\xi $. In stark contrast, the correlations in Eq.~\eqref{eq:CkmodBSSlinshear} extend beyond $\xi$, decaying algebraically. Accordingly, shear is a qualitative correction, in contrast to the above findings for model A. In order to illustrate this further, we consider the limit of small $\xit=\xi/\modx$, i.e., $\xit\ll 1$:
\al{
\dot\gamma C^{(1)}(\bm x,t\!\to\!\infty)\overset{\xit\to0}{\longrightarrow} k_BT\frac{3 \Omega^{(\bm x)} _x \Omega^{(\bm x)} _y}{2 \pi m \lamt ^2 \modx ^3}\xit^{2}.\label{eq:CmodB}
} 
The latter result exhibits again the aforementioned  difference to Eq.~\eqref{eq:CkmodASSx} in that it is scale-free with respect to $\modx$. However, it is also structurally distinct from Eq.~\eqref{eq:CPRL} in that it carries $\xi$ as a prefactor.

The expression in Eq.~\eqref{eq:CkmodBSSlinshear} diverges for $\xi\!\to\!\infty$, which in this limit points to a non-analytic dependence on $\dot\gamma$~\cite{dhont}. Indeed, Eq.~\eqref{eq:linresplengths} requires ever smaller values of $\dot\gamma$ in order for Eq.~\eqref{eq:CkmodBSSlinshear} to remain the leading term. This is in contrast to model A, for which $\lambda$ increases with $\xi$.

\section{Quench and shear}
In Secs.~\ref{sec:review} and \ref{sec:SS} the effects of quenching and shearing were discussed separately. Here, we shall analyze their combined effects. Since in model A no post-quench LCRs are found, we restrict our studies to model B throughout. As before, quantities before the quench are denoted with subscript $I$, and parameters after the quench carry no subscript.

\label{sec:modBDyn}
\subsection{Formal solution}
\label{sec:formalsol}

The formal solution given in Eq.~\eqref{eq:Cktsol} yields
\al{
C(\bm k, t)& = \Xi[\bm k_{\dot\gamma}(t)]C(\bm k_{\dot\gamma}(t),0;T_I) \nl
&\quad + 2\mu_B k_B T\int_0^{t} ds\; \Xi[\bm k_{\dot\gamma}(t-s)]k^2_{\dot\gamma}(t-s), 
\label{eq:CktModB}
}
where we have introduced the advected wave-vector
\al{
\bm k_{\dot\gamma}(t) = (k_x,k_y+2t\dot\gamma k_x, k_z)
}
and the function
\al{
&\Xi[\bm k_{\dot\gamma}(t)] =   \nl
&\exp\Big\{
-2 \mu_B t
\big[
k^2(\kappa k^2 + m) +2t \dot \gamma  k_xk_y \left(m+2 \kappa  k^2\right)\nl
& \qquad \quad  + \frac 4 3 k_x^2 t^2 \dot\gamma^2 (m +2 \kappa (k^2 + 2 k_y^2) )
+8 t^3 \dot\gamma ^3 \kappa  k_x^3 k_y  \nl
&\qquad \quad+\frac{16}{5} t^4 \dot \gamma ^4 \kappa k_x^4
\big]
\Big\}.
\label{eq:Xiformal}
}
The first term on the rhs of \Eref{eq:CktModB} represents the relaxation of the initial equilibrium correlations, with $C(\bm k, 0;T_I) =  {k_B T_I}/[{\kappa_I k^2 + m_I}]$. 
In the absence of  a closed solution of the Fourier inversion of Eq.~\eqref{eq:CktModB}, we shall analyze this expression in various limits. For $\modx/\xi = \infty$, Eq.~\eqref{eq:CktModB} can be inverted analytically, as discussed in Sec.~\ref{sec:modBDynLocal}.  
Subsequently, we shall provide perturbative expressions for finite but large values of $\modx/\xi$ in Sec.~\ref{sec:modBDynNonLoc}.

\subsection{Explicit solution in the limit \texorpdfstring{$|\bm x|/ \xi = \infty$}{}}
\label{sec:modBDynLocal}

\subsubsection{Correlations between spatially fixed points}

\begin{figure}[t]
\centering
\includegraphics[width=0.99\columnwidth]{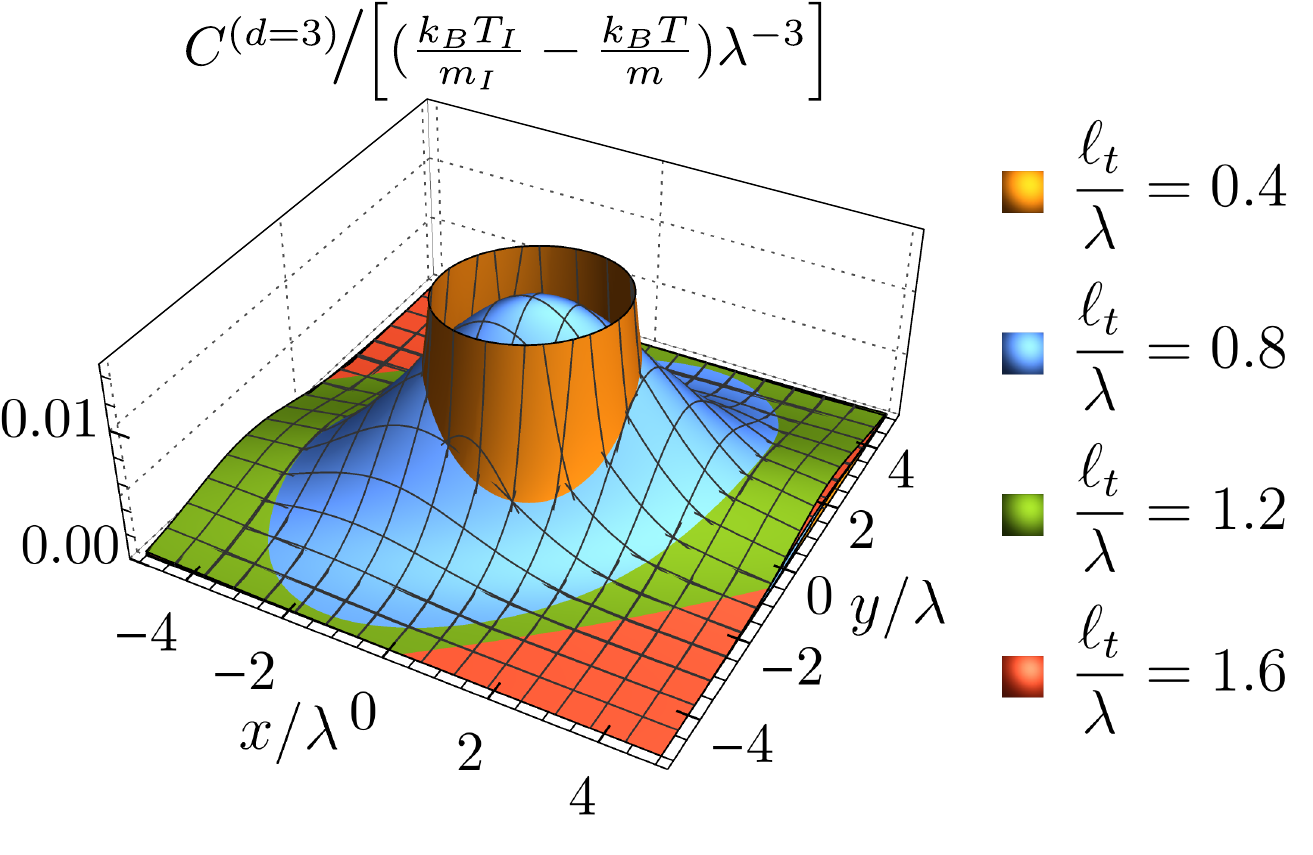}
\caption{The post-quench correlation function from \Eref{eq:CqxtModBLocal} [see, in particular, the 2nd line] in units of $({k_B T_I }/{m_I} -{k_B T}/{m})/\lambda^3$ for $d=3$ and $z=0$, i.e., $\bm x^\perp = (y,z\!=\!0)$ [see also \Eref{eq:CxtModBLoc}]. Since $\ell_t= \sqrt{D t}/\modx$ and $\lambda = \sqrt{D/\dot\gamma}$, the four different surfaces represent different values of $\sqrt{\dot\gamma t}=\ell_t/\lambda$. At later times (i.e., larger values of $\ell_t$), the amplitude of the correlation function is stretched by shear (relative to the symmetric form at early times). The vertical axis is truncated, because the limit $\lt\to0$ is (infinitely) sharply peaked.}
\label{fig:CQlocal}
\end{figure}

In the limit of a large observation length scale relative to $\xi$, i.e., $|\bm x|/ \xi\to \infty$, the terms $k^2 \kappa$ in the time integral in \Eref{eq:CktModB} [see also \Eref{eq:Xiformal}] can be dropped, and the integral can be carried out explicitly. We find  (with $D=\mu_B m$) 
\al{
\!C(\bm k, t) =&\frac{ k_B T}{m}  + \left( \frac{k_B T_I }{m_I} -\frac{k_B T}{m} \right) \times\nl
&\exp\left[-2 D t (k^2+ 2t\dot{\gamma}  k_xk_y+\frac 4 3 t^2{\dot{\gamma}} ^2 k_x^2 )\right].
\label{eq:CktlocalT0}
}
\begin{widetext}
This expression can be Fourier-inverted, yielding a result which is valid in any order of the shear rate (for $|\bm x|/ \xi\to \infty$, we drop the first term in Eq.~\eqref{eq:CktlocalT0}, 
which amounts to a local contribution $\propto \delta^{(d)}(\bm x)$):
 \al{
C(\bm x, t) &= \left( \frac{k_B T_I }{m_I} -\frac{k_B T}{m} \right)
\frac{\exp 
\left(
-\frac{3 |\bm x|^2 - 6 \dot{{\gamma}}  t x y + \dot{{\gamma}} ^2 t^2 \big[3y^2+\modsq{\bm x^\perp}\big]}{24 D t+8 \dot{{\gamma}} ^2 D t^3}
\right)}
{(8\pi D t)^{d/2} \sqrt{1+\dot{{\gamma}} ^2 t^2/3}}.
\label{eq:CxtModBLoc}
} 
Here $\bm x^\perp$ is the component of $\bx$ perpendicular to the flow direction $\bm e_x$. We note that $C$ has precisely the functional form of the probability density of a particle diffusing in shear flow [compare, e.g., Ref.~\cite{dhont}]. Equation~\eqref{eq:CxtModBLoc} differs, however, in that here the time is twice as large. This is because $C$ follows Eq.~\eqref{eq:dtCk}, which carries an extra factor of two compared to the diffusion equation. (This is a generic observation when comparing dynamics of a stochastic variable and its correlation function.) The prefactor in \Eref{eq:CxtModBLoc} shows that this long-ranged contribution is absent without a quench. 
$ C(\bm x,t)$ in Eq.~\eqref{eq:CxtModBLoc} illustrates how the quench-induced correlations provided in \Eref{eq:CPRL} are distorted by shear. We rewrite \Eref{eq:CxtModBLoc} in terms of the time scale $t^* = D t/|\bm x|^2$ [see \Eref{eq:tstardef}], the length scale $\lamt = \lambda/\modx$, and the angular variables $\Omega_\alpha \equiv \Omega_\alpha^{(\bm x)}$ [see \Eref{eq:Omegadef}]:
\al{
\frac{C_{\textrm{}}(\bm x,t)}{{k_B T_I }/{m_I} -{k_B T}/{m}}
&= 
\frac
{\exp\left( -\frac{3-6 \lamt^{-2} t^* \Omega _x \Omega _y+ \lamt^{-4} \left(t^*\right)^2 \left[3 \Omega _y^2+\Omega _\perp^2\right]}
{24 t^*+ 8 \lamt^{-4} \left(t^*\right)^3} \right)}{\modx^d (8\pi t^*)^{d/2}\sqrt{1+\left(t^*\right)^2/3\lamt^4 }}\nl
&= 
\frac
{\exp\left(-\frac{\lambda^2}{\lt^2} \frac{3(x^2+y^2)/\lambda^2 + (xy/\lambda^2)(\lt^2/\lambda^2)+ (3y^2+\modsq{\bm x^\perp})/\lambda^2 (\lt^4/\lambda^4)}{3+\lt^4/\lambda^4}  \right)}{\lambda^d (8\pi)^{d/2}(\lambda/\lt)^d\sqrt{1+\lt^4/3\lambda^4 }} \nl
&\simeq 
\begin{cases}
\frac{e^{-\frac{1}{8 t^*}}}{(8\pi t^*)^{d/2}}\Big[ \frac{1}{\modx^d }+\frac{\Omega _x \Omega _y}{4\modx^{d}}\lamt^{-2} 
+ \frac{\left(\Omega_x^2-4 t^*\right) \left(4 t^*+3 \Omega_y^2\right)}{1536 \sqrt{2} \modx^d\pi ^{3/2}  \left(t^*\right)^{3/2}}\lamt^{-4}\Big],\;\lambda\gg\{\lt,\modx\},\\
\frac{e^{-\frac{3 \Omega _y^2+\Omega _\perp^2}{8 t^*}}}{(8\pi t^*)^{d/2}}\Big[ \frac{\sqrt{3} }{\modx^{d}t^*}\lamt ^2 +\frac{3 \sqrt{3}   \Omega _x \Omega _y}{4 \modx^{d} \left(t^*\right)^3} \lamt ^4 \Big],\;\lambda\ll\{\lt, \modx\}.
\end{cases} 
\label{eq:CqxtModBLocal}
}
Here, in the final step,  we have expanded the expression for $\lambda\gg\{\lt,\modx\} $ (i.e., linear response in $\dot\gamma$) and $\lambda\ll \{\lt, \modx\}$ (strong shear), and used $\Omega_\perp^2 = \Omega_y^2+\Omega_z^2$ in $d=3$.  At zero shear ($\dot\gamma=0$), the correlations of \Eref{eq:CPRL} are recovered. In steady state, $C(\bm x, t\!\to\!\infty)$ vanishes. Equation \eref{eq:CqxtModBLocal} shows that the correlation between two points after a quench depends on the orientation of the vector connecting them (relative to the shear velocity $\bm v$). 
Equation~\eref{eq:CqxtModBLocal} is illustrated in Fig.~\ref{fig:CQlocal} for certain choices of parameters, where the functional form in the second line of \Eref{eq:CqxtModBLocal} was used.
\end{widetext}

\subsubsection{Correlations in the frame co-moving with shear}
\label{sec:modBDynComov}

\begin{figure*}[t]
\centering 
\includegraphics[width=0.484\textwidth]{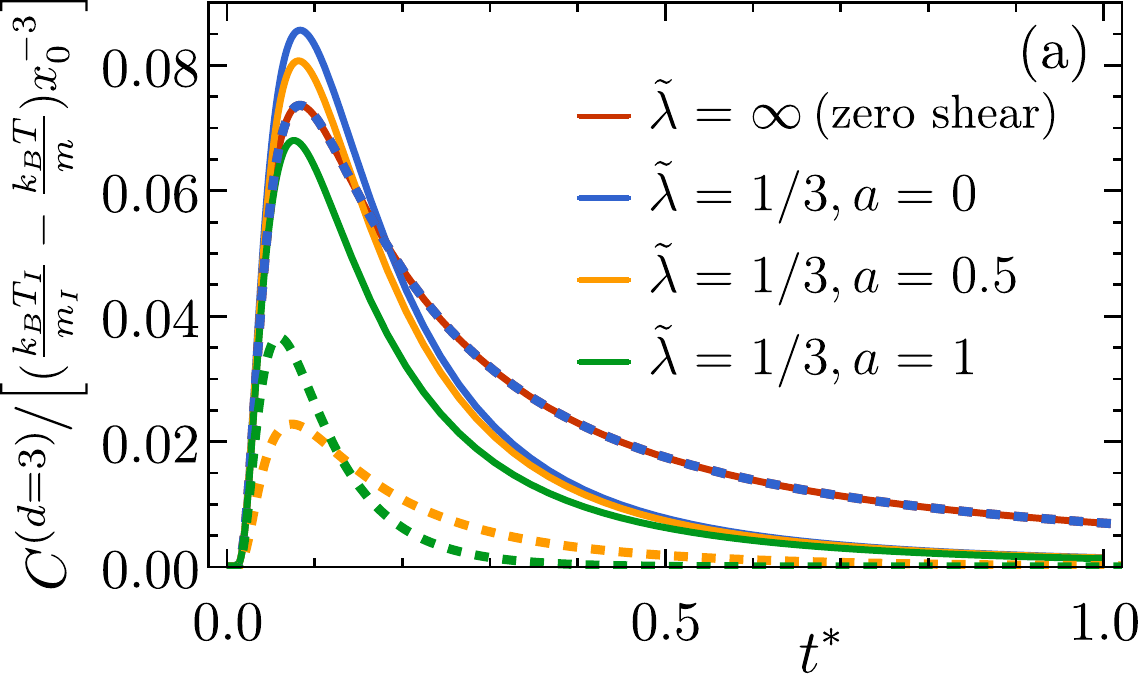}
\hspace{0.4cm}
\includegraphics[width=0.48\textwidth]{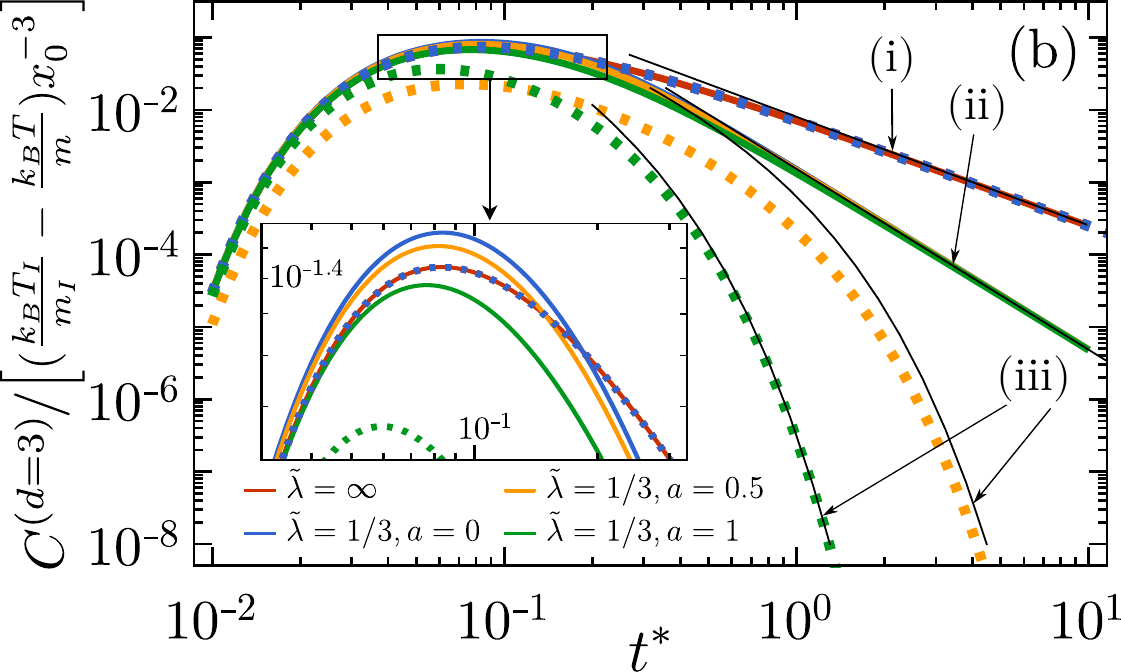}

\caption{
The post-quench correlation function between two points following co-moving trajectories [i.e., their connecting vector is $\bm x(t;\dot\gamma)\! = \!\bm x_0 + y_0\dot\gamma t \bm e_x$], in units of $({k_B T_I }/{m_I} -{k_B T}/{m})/|\bm x_0|^3$. Panel (a) [(b)] reports the results in linear [logarithmic] scales. Solid lines represent Eq.~\eref{eq:CQcomov} in the case that the medium is sheared, whereas dashed curves correspond to two points following the same trajectory $\bm x(t)$, but in an unsheared medium [obtained by setting $\dot\gamma=0$ in \Eref{eq:CxtModBLoc}, and evaluating the result at $\bm x\! = \!\bm x(t;\dot\gamma)$]. Thus both the solid and the dashed curves depend on $\lamt\!= \!\lambda/|\bm x_0|$ with $\lambda\! =\! \sqrt{D/\dot\gamma}$, but in different ways. For $\lamt\!=\infty$ (zero shear), the solid and the dashed lines coincide, as both reduce to the case of spatially fixed points in an unsheared medium [\Eref{eq:CPRL}]. For $\lamt\!\neq\!\infty$, the maximum of $C$ depends on $a\!=\! \frac{|\bm x_0^\perp|}{|\bm x_0|}\in[0,1]$. For $a\!=\!0$, the dashed curves correspond to the case with $\lamt \!= \!\infty$, as $\bm x(t) \!= \!\bm x_0$, $\forall t$, for $\bm x_0^\perp \!=\!\bm 0$. In panel (b), labelled arrows and thin black curves show the late-time asymptotes (i) $\frac 1 {16\sqrt 2 (\pi t^*)^{3/2}}$, (ii) $\frac {\sqrt{3/2}\lamt^2} {16 \pi^{3/2}(t^*)^{5/2}}$, and (iii) $\frac {\exp(-a^2 t^*/8\lamt^2)} {16 \sqrt 2(\pi t^*)^{3/2}}$. All solid curves for $\lamt\neq\infty$ decay asymptotically as  $\propto(t^*)^{-5/2}$; differences at early times are shown in the inset. For further details see main text after \Eref{eq:CQcomov}.
}
\label{fig:Ccomov}
\end{figure*}

It is instructive to consider two points which are advected by shear flow, i.e., separated by the vector $\bm x(t) = \bm x_0 + \bm v t =  \bm x_0 + y_0\dot\gamma t \bm e_x$, because this is the natural trajectory of a particle in flow. The corresponding post-quench correlations in the co-moving frame (as indicated by subscript ``c-m'') can be inferred from Eq.~\eref{eq:CxtModBLoc}; for $d=3$ we find
\al{
\frac{C_{\textrm{c-m}}^{(d=3)}}{\frac{k_B T_I }{m_I} -\frac{k_B T}{m} } &= \frac{\exp \left(-\frac{|\bm x_0|^2 + \dot \gamma ^2 t^2 |\bm x_0^\perp|^2/3}{8 D t \left(1+\dot\gamma ^2 t^2/3\right)}\right)}{(8\pi D t)^{3/2}\sqrt{1+\dot\gamma ^2 t^2/3} }\nl
&= \frac{ \exp\left(-\frac{1 + a (t^*)^2/3\lamt^4}{8t^*[1+ (t^*)^2/3\lamt^4]}\right)}{(8\pi t^*)^{3/2} |\bm x_0|^3 \sqrt{ 1+ (t^*)^2/3\lamt^4 }}.
\label{eq:CQcomov}
}
Here $\bm x_0^\perp = (y_0,z_0)$ labels the components of $\bm x_0$ perpendicular to the shear direction $\bm e_x$, with $a \equiv |\bm x_0^\perp|/|\bm x_0|$. In the last line, $t^*$ is an abbreviation for $t^*_{|\bm x_0|}= D t/ |\bm x_0|^2$ [see \Eref{eq:tstardef}], i.e., we rescale time by the time scale of diffusion across the {initial separation} $|\bm x_0|\neq0$.

Figure~\ref{fig:Ccomov} compares \Eref{eq:CQcomov} with the following expressions: \\
\textbf{(i)}~The two-point correlation function of Eq.~\eqref{eq:CPRL}, i.e., for the system without shear and evaluated at a distance $|\bx_0|$ between the two points, as given by the curve with $\lamt=\infty$. We note that, especially at early times, the correlations of \Eref{eq:CQcomov} can be larger than the ones of the corresponding quiescent system, even though the correlations are taken between points at larger distances. The maximum of the curve can be tuned by choosing $a = \frac{|\bm x_0^\perp|}{|\bm x_0|}\in[0,1]$. However, at late times shearing speeds up the decay of correlations. While equilibrium correlations decay as $ (t^*)^{-3/2}$, the curve corresponding to \Eref{eq:CQcomov} decays as $(t^*)^{-5/2}$ [see Fig.~\ref{fig:Ccomov} (b)]. \\
\textbf{(ii)}~The correlation function of an \textit{unsheared} fluid after a quench [see Eq.~\eqref{eq:CPRL}] evaluated at the distance $|\bm x(t)|$. This result, which still depends on $\dot\gamma$ (and thus $\lamt$) via $\bm x(t)$, illustrates the effect of moving the observation points along a ``shear trajectory'', in contrast to shearing the medium itself. These correlations are in general weaker than those resulting from \Eref{eq:CQcomov}, and exhibit a qualitatively different behavior in that they decay exponentially at late times. The latter occurs because, for any finite $a$, the two points move apart faster than the diffusion of the correlations. For $a\to0$, the correlations between co-moving points in the unsheared system collapse onto the curve corresponding to $\lamt = \infty$, because in this limit $\bm x(t) \to \bm x_0\;\forall t$, i.e., the stationary case is recovered. Generically, correlations are maximal for $a=0$ both in sheared and unsheared systems.

\subsection{Perturbative solution for non-zero \texorpdfstring{$|\bm x|/\xi$}{} with weak shear}
\label{sec:modBDynNonLoc}
In this subsection we evaluate $C(\bm k, t)$ from Eq.~\eref{eq:CktModB} in order to include effects of a nonzero correlation length $\xi$, especially regarding the change of $\xi$ during a quench. The expression in Eq.~\eref{eq:CktModB} can be determined analytically in the limit of small shear and large  $\modx/\xi$. For $d=3$ we find 
\begin{widetext}
\al{
C^{(d=3)}(\bm x, t)= &\frac{1}{ |\bm x|^3} \left( \frac{k_B T_I }{m_I} -\frac{k_B T}{m} \right)
\left[
h^Q_{0,0}(t^*) 
+\tilde \xi ^2h^Q_{0,\xi^{2}}(t^*)
+ \frac{\Omega_x \Omega_y}{\tilde\lambda^2} 
h^Q_{\dot\gamma,0}(t^*) 
+ \frac{\Omega_x \Omega_y \xit ^2}{\lamt^2} 
h^Q_{\dot\gamma,\xi^{2}}(t^*) \right]\nl
&+ \frac{k_B T}{m |\bm x|^3} \left[ \frac{\Omega_x \Omega_y\xit ^2}{\lamt^2} 
h^S_{\dot\gamma,\xi^{2}}(t^*) \right]
+ \frac{k_B T_I}{m_I |\bm x|^3} \Delta \tilde\xi^2
\left[ h^R_{0,\Delta\xi^2}(t^*)
+ \frac{\Omega_x \Omega_y}{\lamt^2} 
h^R_{\dot\gamma,\Delta\xi^2}(t^*) \right]\nl
&+ \mathcal{O}\left(\xit^4,\Delta\xi^4,\lambda^{-4}\right),
\label{eq:C(t)ModBCombined} 
}

\end{widetext}
where
\begin{subequations}\label{eq:A(t)ModBCombined}
\begin{align}
h^Q_{0,0}(t^*)&= \frac{e^{-\frac{1}{8 t^*}}}{(8\pi t^*)^{3/2}}=4 h^Q_{\dot\gamma,0}(t^*),\nln
h^Q_{0,\xi^{2}}(t^*)&= \frac{e^{-\frac{1}{8 t^*}} (48 (1-7 t^*) t^*-1)}{128(8\pi )^{3/2} (t^*)^{9/2}} ,\nln
h^Q_{\dot\gamma,\xi^{2}}(t^*)&= \frac{e^{-\frac{1}{8 t^*}} (16 (4-41 t^*) t^*-1)}{512(8\pi )^{3/2}(t^*)^{9/2}}, \nln
h^S_{\dot\gamma,\xi^{2}}(t^*)&= \frac{3}{2\pi} \text{erfc}\left(\frac{1}{2 \sqrt{2} \sqrt{t^*}}\right) 
+\frac{e^{-\frac{1}{8 t^*}} \left(12 t^*+1\right)}{2(8\pi )^{3/2}\left(t^*\right)^{3/2}}, \nln
h^R_{0,\Delta\xi^2}(t^*)&= \frac{e^{-\frac{1}{8 t^*}} \left(12 t^*-1\right)}{(8 \pi )^{3/2} 16 \left(t^*\right)^{7/2}}=4h^R_{\dot\gamma,\Delta\xi^2}(t^*).
\end{align}
\end{subequations}
Above, $t^* = t^*_{\modx}$ [see \Eref{eq:tstardef}], $\Delta\xi = \xi-\xi_I$ reflects any change in correlation length during the quench, and $\Delta\xi^n \equiv \xi^n-\xi_I^n$. Further, we use $\tilde l = l/\modx$ for $l \in \{\xi,\lambda,\Delta\xi\}$.

Equations~\eref{eq:C(t)ModBCombined} and \eref{eq:A(t)ModBCombined} reveal explicitly the origin of the various contributions due to quenching and shearing. The first line of Eq.~\eqref{eq:C(t)ModBCombined} is generated by a change of the ratio $k_BT/m$ during the quench (denoted by the superscript ``$Q$''), expanded in terms of small shear and a large length ratio $\modx/\xi$. This contribution mostly aligns with the discussion in Sec.~\ref{sec:modBDynLocal}, extended by including finite values of $\modx/\xi$.

The first term in the second line is the only one with a nonzero long-time value. This term describes a system with shear (denoted by a superscript ``$S$'') starting at $t=0$ in the absence of a quench. It thus relaxes to the result of Eq.~\eqref{eq:CmodB}, with $\lim_{t^* \to \infty} h^S_{\dot\gamma,\xi^{2}}(t^*) = \frac{3}{2\pi}$.  The second term in the second line represents relaxation (denoted by a superscript ``$R$'') after a quench of the correlation length $\xi$ itself, as captured by $\Delta\xi$. Concerning the nomenclature for the functions $h$, subscripts denote the order of perturbation in shear and correlation length, respectively.

\begin{figure*}[t]
\centering
\includegraphics[width=0.48\textwidth]{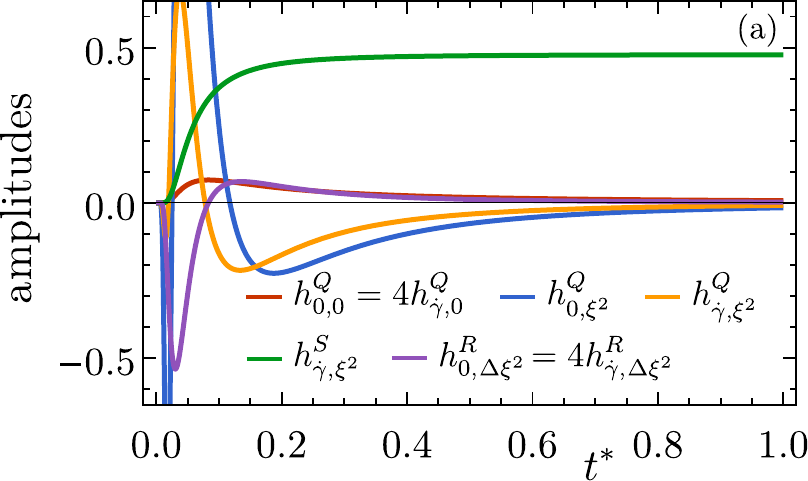}
\hspace{0.4cm}
\includegraphics[width=0.474\textwidth]{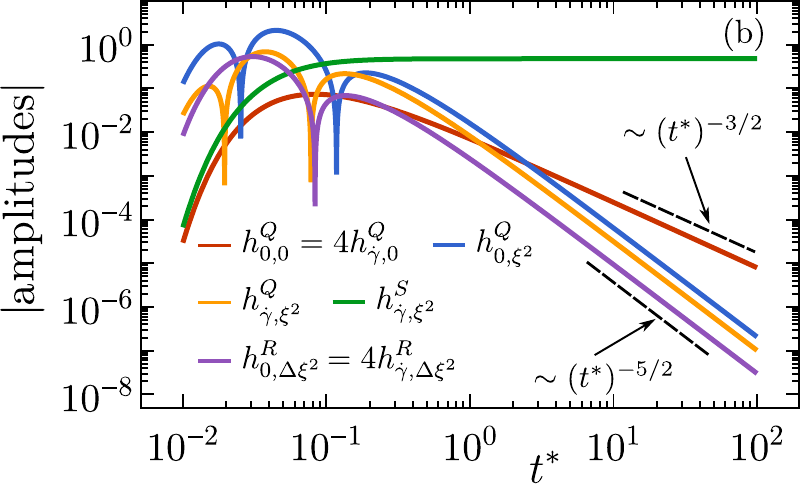}
%
\caption{Model B with quench and shear: dimensionless amplitudes [Eq.~\eref{eq:A(t)ModBCombined}] of the various contributions to the correlation function given in Eq.~\eref{eq:C(t)ModBCombined}, displayed on linear (a) and logarithmic (b) scales as functions of $t^* = D t/\modx^2$. The contributions with superscript $Q$ arise from quenching $k_B T/m$; those with superscript $R$ come about upon quenching $\xi$; $h^S_{\dot\gamma,\xi^2}$ captures the distortion of small inherent correlations ($\xi\!\neq\!0$) by shear. For further details see the main text next to Eqs.~(\ref{eq:C(t)ModBCombined}) and (\ref{eq:A(t)ModBCombined}).
}
\label{fig:A(t)ModB}
\end{figure*}

Figure \ref{fig:A(t)ModB} shows the time dependence of the various amplitudes. Panel (a) provides linear scales, while panel (b) displays the long-time behavior on logarithmic scales. We see that the steady-state contribution $h^S_{\dot\gamma,\xi^{2}}(t^*)$ is approached algebraically at late times. All shear-dependent contributions exhibit a dependence on the orientation of $\bx$.

Finally we point out the difference between quenching the ratio $k_BT/m$ and quenching the correlation length $\xi$ (under the proviso that in an experiment these quantities can be quenched independently): at leading order (in $t^*$), a quench of $\xi$ only renders correlations which decay more rapidly in time than the corresponding correlations due to quenching the ratio $k_BT/m$. (This is easily inferred from the exponents of the algebraic long-time tails in Fig.~\ref{fig:A(t)ModB}.) The spatial algebraic prefactors of the various contributions in \Eref{eq:C(t)ModBCombined} also differ, so that (for fixed $\xi$ and $\lambda$), the contributions $\propto h^{Q}_{0,\xi^{2}}$ and $h^{Q/R}_{\dot\gamma,0}$ are shorter-ranged than those $\propto h^{Q}_{0,0}$ and $h^{Q/S/R}_{\dot\gamma,\xi^2}$.

\section{Non-equilibrium forces between two small inclusions in a shear field}
\label{sec:incsec}

Long-ranged correlations give rise to fluctuation-induced forces between objects which confine the fluctuations~\cite{kardargolestanian1999}. Such forces occurring after a quench have been analyzed, for instance, in Refs.~\cite{gambassi2008EPJB,deangopinathan2009JStatMech,deangopinathan2010pre,rohwer2017transient,rohwer2018forces}. In the spirit of the above discussions, we now investigate post-quench forces between inclusions in shear. We start by generalizing  the  formalism  of Ref.~\cite{deangopinathan2010pre} for the computation of time-dependent non-equilibrium forces after a quench with shear, and then apply these results to the case of small inclusions,  using the correlation functions computed above.

\subsection{Nonequilibrium forces with quench and shear: extending the framework of Ref.~\cite{deangopinathan2010pre}}
\subsubsection{General derivation}
\label{sec:GenDeriv}

\begin{figure}[t]
\centering
\includegraphics[width=0.6\columnwidth]{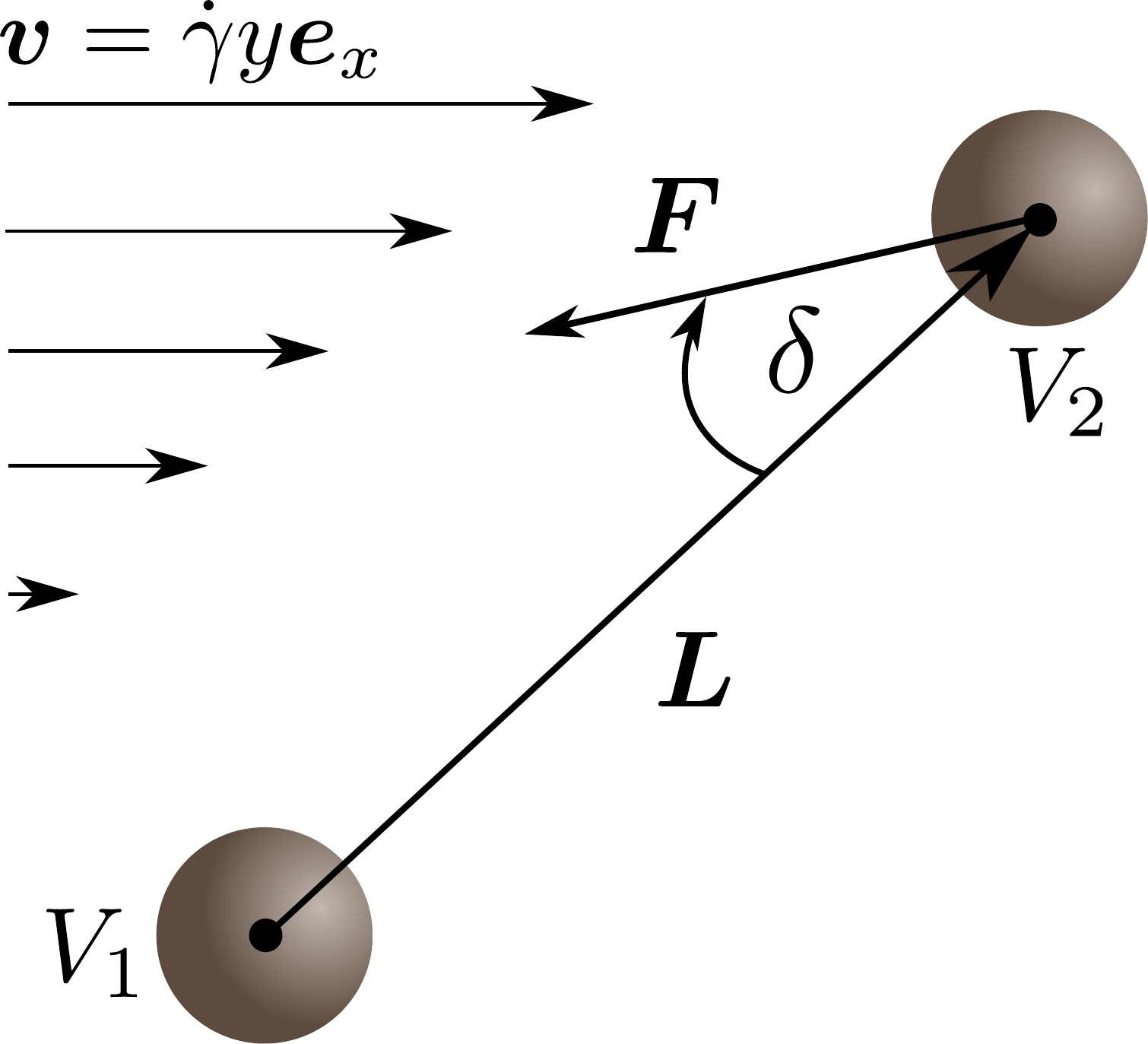}
\caption{Two inclusions (volumes $V_{1,2}$) immersed in a correlated fluid, separated by a vector $\bm L$. For inclusions held at a fixed relative position, one has $\bm L = \bm L_0$. For inclusions following a co-moving (advected) trajectory in shear flow, $\bm L = \bm L(t) \equiv \bm L_0 + \dot\gamma t L_{0,y} \bm e_x$. The separation of the inclusions is taken to be much larger than their radii. In this limit, $\bm L$ becomes independent of the choice of reference points within $V_i$ to be connected by $\bm L$.
}
\label{fig:IncSketch}
\end{figure}

In the context of model A and model B dynamics, Dean and Gopinathan derived a formalism for computing non-equilibrium forces which emerge between immersed objects after a quench in systems described by bilinear Hamiltonians of the form \cite{deangopinathan2010pre}
\begin{align}
H = \frac 1 2 \int \textrm d^dx \,\textrm d^d x' \;\phi(\bx)\Delta(\bx,\bx',\bm L)\phi(\bx').
\end{align}
The force $\ave{\bm F(t)}=-\ave{\nabla_{\bm L}H[\phi(t)]}$ (averaged over noise realizations) is computed from the \textit{instantaneous} configuration of $\phi$, and $\bm L$ is the relevant vector separating the external objects (e.g., two plates or two finite-sized inclusions). In Appendix \ref{sec:deanappend}, we extend this formalism in order to include shear flow. The main result is that the Laplace transform of the time-dependent (non-equilibrium) force following a quench can be computed from an \textit{effective equilibrium theory}:
\al{
\ave{\bm F(s)} = \frac{k_B T}{s} \nabla_{\bm L} \ln[Z(\Delta_s^{(\dot\gamma)})],
\label{eq:LF(s)}
}
where $f(s) = \mathcal L [f(t)](s)=\int_0^\infty dt \;e^{-t s} f(t)$ denotes the Laplace transform of $f(t)$. Equation~\eqref{eq:LF(s)} states that the non-equilibrium force is given by the equilibrium force corresponding to an $s$-dependent Hamiltonian with $\Delta_s^{(\dot\gamma)} = \Delta + s(R^{(\dot\gamma)})^{-1}/2$, where $R^{(\dot\gamma)} = R+S\Delta^{-1}$.  Here,  $R(\bx,\bx') = \hat\mu_{A/B}(\bx)\delta(\bx - \bx')$ and $S(\bx,\bx') = \dot\gamma y\partial_x \delta(\bx - \bx ')$ corresponds to the advection term in the Langevin equation \eref{eq:Langevin}. This $s$-dependent Hamiltonian leads to the following $s$-dependent partition sum:
\begin{align}
Z(\Delta_s^{(\dot\gamma)}) = \!\int\! \mathcal D \phi \,e^{-\frac \beta 2 \int\! \textrm d^dx \,\textrm d^d x'\phi(\bx)\Delta_s^{(\dot\gamma)}(\bx,\bx',L)\phi(\bx')}.
\end{align}
Accordingly, the force is obtained by taking the gradient with respect to the separation $\bm L$, as stated in Eq.~\eqref{eq:LF(s)}.

We emphasize that Eq.~\eqref{eq:LF(s)} rests on the assumption that the external objects do not alter the shear flow. This is expected to be valid in the case of plates oriented parallel to shear, or for the small inclusions which will be investigated below.

Remarkably, at times long after the quench, the force in Eq.~\eqref{eq:LF(s)} adopts exactly the equilibrium form, 
\al{
\bm F(t\!\to\!\infty) = k_B T \nabla_{\bm L} \lim_{s\to0}\ln[Z(\Delta_s^{(\dot\gamma)})],
\label{eq:LF(tti)}
}
but with a shear-dependent pseudo-partition sum. In \Eref{eq:LF(tti)} and in what follows, we suppress the implied average for brevity [compare \Eref{eq:LF(s)}].

\subsubsection{Two inclusions of finite size}
We now apply this result to determine non-equilibrium forces between two inclusions with volumes $V_1$ and $V_2$, respectively, separated by a vector $\bm L$ pointing from the first to the second inclusion, in the limit of large separation ($L\gg V_i^{1/d}$). We model these \textit{inc}lusions in terms of local, quadratic contributions to the Hamiltonian:
\al{
H=\int \textrm d^dx \left[\frac \kappa 2 (\nabla\phi)^2+\frac m 2 \phi^2(\bm x)\right]+H_{\textrm{inc}},
\label{Htot}
}
with the first (bulk) term given in Eq.~\eqref{eq:H}, and where
\al{
H_{\textrm{inc}} = \frac{c_1}{2} \int_{V_1} \textrm d^dx \, \phi(\bm x)^2 + \frac{c_2}{2} \int_{V_2} \textrm d^dx \, \phi(\bm x)^2
\label{Hinc}
}
represents the inclusions in terms of coupling constants $c_1$ and $c_2$. Accordingly, the above formalism can be applied. Thus the inclusions are modelled by a contrast of the mass inside the inclusions relative to the bulk value $m$:
\al{
m(\bm x) = 
\begin{cases}
m + c_i, & \bm x \in V_i,\\
m, &\mbox{elsewhere}.
\end{cases}
\label{eq_mx}
}
For simplicity, we consider $k_BT_I/m_I=0$, i.e., there are no fluctuations before the quench (corresponding, e.g., to a low initial temperature). The quench gives rise to a non-equilibrium force which can be expressed in terms of a pseudo-potential $\mathcal V$, derived from a cumulant expansion of $H_{\textrm{tot}}$ (see Appendix \ref{sec:deanappend}):
\al{
\mathcal V(\bm L,t) &=  
\frac{k_B T} 2\alpha_1 \alpha_2
\mathcal L_{s\to t}^{-1}  \Big[\frac 1 s \big(
C(\bm L,s) \frac{s \;m}{k_B T}
\big)^2\Big].
\label{eq:VLt}
}
Here $C(\bm L,s) = \mathcal L [C(\bm L,t)](s)$ is the Laplace transform of the equal-time two-point correlation function in the bulk, at separation $\bm L$ [see \Eref{eq:Cs}], and $\alpha_i = V_i c_i / m$. The force on the first inclusion is
\al{
\bm F(\bm L,t) = -\nabla_{\bm L} \mathcal V(\bm L,t).
\label{eq:FincStat}
}
In Eq.~\eqref{eq:FincStat}, the displacement vector $\vec{L}$ can depend explicitly also on time, for instance if one considers the force between moving objects, as discussed below. This case will be indicated by $\Lt$, while $\L0$ refers to spatially fixed inclusions.

In the long time limit, one has
\al{
\mathcal V(\bm L,t\!\to\!\infty) &=  
\frac{k_B T} 2\alpha_1 \alpha_2
\left[\frac{m \;C(\bm L,t\!\to\!\infty)}{k_B T}\right]^2
,
\label{eq:VLtti}
}
so that the force in the steady state with shear can be inferred easily from the equal-time two-point correlation function in the bulk, adopting the same form in terms of correlation functions as in equilibrium.

\subsection{Forces in steady state under shear}
\label{sec:FSS}
Using Eq.~\eqref{eq:VLtti} and the results of Sec.~\ref{sec:SS}, one can directly provide the non-equilibrium forces in the steady state under shear. For model A, the correlations in Eq.~\eqref{eq:CkmodASSx} yield the following force vector:
\al{
\bm F_{A}(t\!\to\!\infty)=&\frac{\alpha _1 \alpha _2 k_BT }{256 \pi ^2 |\bm L_0|^{7}} \frac{ e^{-{2}/{\xit^2}}}{ \lamt^4 \xit^6}
\nl
&\times\!
\begin{pmatrix}
\Omega _x \Omega _y^2 \left[\xit^2-\left(\xit^2+2\right) \Omega _x^2\right]\\
\Omega _x^2 \Omega _y \left[\xit^2-\left(\xit^2+2\right) \Omega _y^2\right] \\
-\left(\xit^2+2\right) \Omega _x^2 \Omega _y^2 \Omega _z
\end{pmatrix}.
\label{eq:FincSSmodA}
}
Here $\L0$ denotes the (stationary) vector joining the inclusions [compare $\bm L(t)$ in Subsec. \ref{sec:Fcomovsec}), the orientation of which is captured by $\Omega_\alpha\equiv\Omega_\alpha^{(\L0)}$; see \Eref{eq:Omegadef}]. $\bm F_{A}(t\!\to\!\infty)$ decays exponentially for $\xit\to0$. 

For model B, we use \Eref{eq:CmodB} in order to obtain the steady state force between the inclusions to leading order in shear:
\al{
\! \bm F_{\textrm B}(t\!\to\!\infty)=\frac{9\alpha _1 \alpha _2 k_BT }{4\pi ^2 |\bm L_0|^{7}}\frac{\xit^4}{\lamt^{4}}
\begin{pmatrix}
\Omega _y^2 \Omega _x \! \left(1-5 \Omega _x^2\right)  \\
\Omega _x^2 \Omega _y \! \left(1-5 \Omega _y^2\right) \\
-5 \Omega _x^2 \Omega _y^2 \Omega _z
\end{pmatrix}\!.
\label{eq:FincSS}
}
Unlike $\bm F_{A}$,  $\bm F_{B}(\xit\to0)$ vanishes according to a power law. Thus, in model B, shear gives rise to a qualitatively relevant steady state force between the inclusions even for small correlation lengths. Strikingly, and in stark contrast to equilibrium steady states, the conservation laws of the underlying dynamics strongly influence the phenomena observed in shear-induced steady states. We also note that, in general, neither $\bm F_{A}$ nor $\bm F_{B}$ are parallel to $\bm L_0$; this also differs from the equilibrium case where, by symmetry, the forces are necessarily along the separation vector~\cite{deangopinathan2010pre,rohwer2017transient}.

\begin{widetext}
\subsection{Forces after quenches}
\label{sec:Faq}

In this subsection we compute time-dependent forces between the two inclusions after a quench. We focus on the limit of vanishing $\xi$, in which no post-quench correlations are observed within model A. Accordingly, the remaining analysis proceeds in terms of model B dynamics; henceforth, the corresponding subscript ``B'' for the force will be dropped.

\subsubsection{Prerequisites}
In order to compute the time-dependent force after quenching, the Laplace transform of the correlation function is required. For two points in the bulk at large separations compared to the correlation length, i.e., $|\vec{x}|/\xi\gg1$, the Laplace transform of \Eref{eq:CqxtModBLocal} reads, in terms of an expansion for small shear rates:
\al{
C(\bm x, s) &= - \frac{k_BT e^{-\frac{\sqrt{s^*}}{\sqrt{2}}} }{8 \pi  D m  \modx }\Big[
1+\frac{\Omega _x \Omega _y}{4 \tilde{\lambda }^2}
+ \frac{3 \sqrt{2} \left(s^*\right)^{3/2} \Omega _x^2 \Omega _y^2+2 s^* \left(\Omega _x^2-3 \Omega _y^2\right)-2 \sqrt{2s^*}-4}{96 \sqrt{2} \left(s^*\right)^{3/2} \tilde{\lambda }^4}
\Big] +\mathcal O (\tilde\lambda^{-6}),
\label{eq:Cxs}
}
where $s^* = \modx^2 s/D$ is the rescaled, dimensionless Laplace variable, and $\Omega_\alpha\equiv\Omega_\alpha^{(\bx)}$ [see \Eref{eq:Omegadef}]. Using Eqs.~\eqref{eq:VLt}, \eref{eq:FincStat}, and \eqref{eq:Cxs},
the force is expanded in terms of powers $n$ of the shear rate:
\al{
\bm F(\bm L,t) &= \sum_{n\geq0}\bm F^{(n)},
\label{eq:FincStatExpand}
}
where the vector $\bm L$ connects the first inclusion to the second one. Equation \eref{eq:FincStatExpand} is obtained by evaluating the relevant correlation functions at $\bm x = \bm L$. First we consider the force between two inclusions which are placed at fixed positions ($\bm L =\L0$). In a second step, we provide the force between two inclusions advected by the shear flow ($\bm L =\bm L(t)$). 
\end{widetext}

\subsubsection{Post-quench force between inclusions at fixed positions}  
\label{sec:PQFfixed}

We now consider the dynamics of the force given in \Eref{eq:FincSS}. In accordance with \Eref{eq:FincStatExpand}, at order $n$, the force between the two (stationary) inclusions is

\begin{figure}[t]
\centering
\includegraphics[width=0.99\columnwidth]{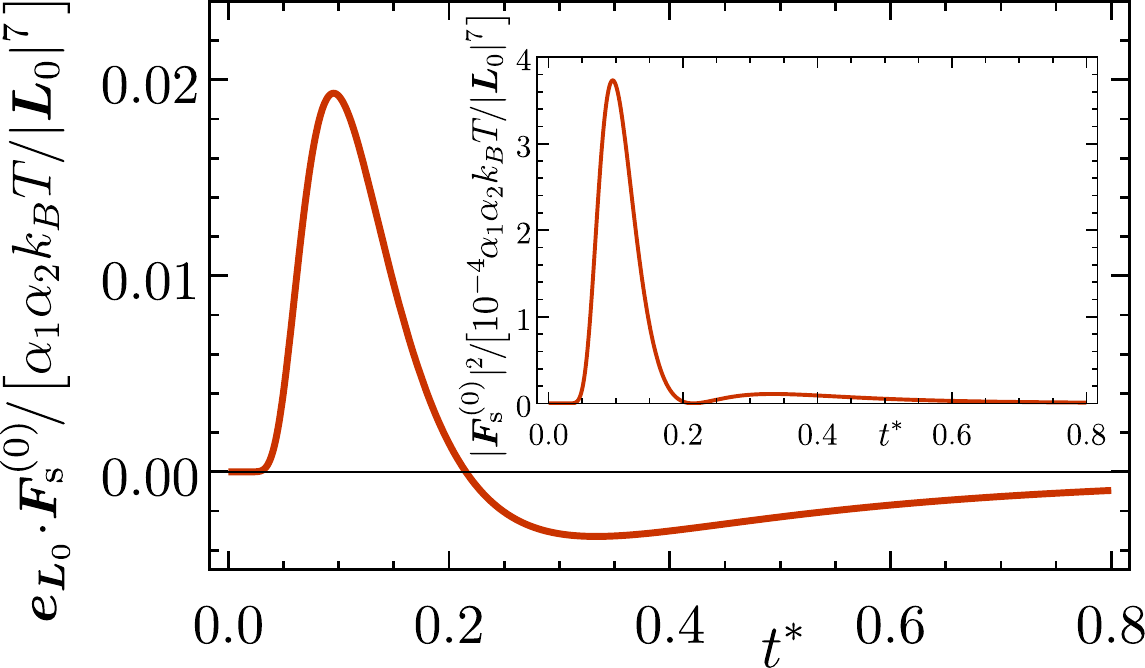}
\caption{The force $\bm F^{(0)}_{\textrm s}(\bm L_0,t)$ between two inclusions following a quench, but in the absence of shear. The inclusions are held at fixed points, separated by the vector $\bm L_0$ [see Eqs.~(\ref{eq:FincStatExpand}-\ref{eq:Fscomponents})]. Here, $t^* = D t/|\L0|^2$ represents the time axis rescaled by the diffusive time scale across the distance $|\L0|$. The force is parallel to the separation vector $\L0$ with unit vector $\bm e_{\L0}=\L0/|\L0|$. Inset: Squared magnitude $|\bm F_{\textrm{s}}^{(0)}|^2$ of the force (in the absence of shear) as function of $t^*$. 
}
\label{fig:f0F0}
\end{figure}

\begin{figure}[t]
\centering
\includegraphics[width=0.99\columnwidth]{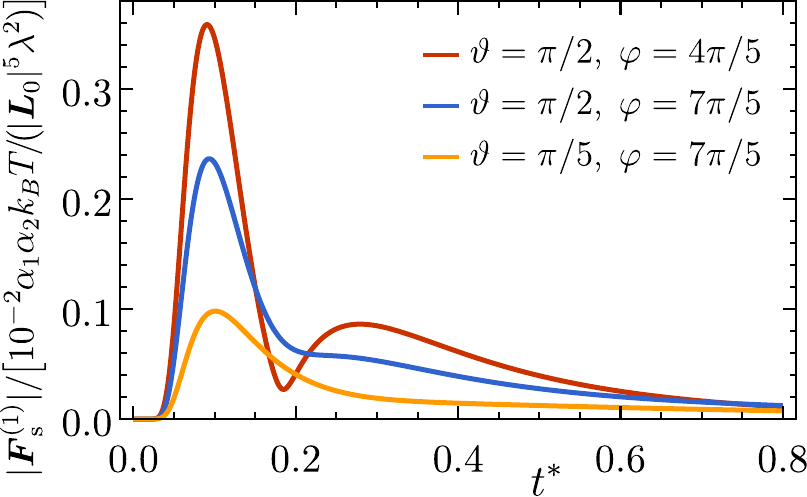}
\caption{The lowest order shear correction $\bm F^{(1)}_{\textrm s}$ [Eqs.~\eref{eq:FincStatExpand2}-\eref{eq:Flongtime}] to the force between two stationary inclusions following a quench. The inclusions are separated by the vector $\bm L_0$, and $t^* = D t/|\L0|^2$. The azimuthal $(\theta)$ and polar ($\varphi$) angles describe the orientation of $\L0$. 
}
\label{fig:F1s}
\end{figure}
\al{
\bm F_{\textrm s}^{(n)}=\frac{\alpha _1 \alpha _2 k_BT}{|\bm L_0|^{7-2n}\lambda^{2n}}  \bm f^{(n)}(t^*).
\label{eq:FincStatExpand2}
}
The subscript ``s'' has been introduced in order to distinguish the present case of stationary inclusions from the co-moving inclusions considered in Subsec.~\ref{sec:Fcomovsec}. Regarding notation, we note that here (and in the following subsections) the vector $\L0$ sets the length scale relative to which we define $\lamt=\lambda/|\L0|$ and the diffusive time $t^*_{|\L0|}=D t/|\L0|^2$ [see \Eref{eq:tstardef}]; henceforth this dependence will not be indicated explicitly. The components $f^{(n)}_\alpha(t^*)$ 
of the vector $\bm f^{(n)} = (f^{(n)}_x,f^{(n)}_y,f^{(n)}_z)$ are dimensionless, time-dependent functions. The first few orders of \Eref{eq:FincStatExpand} can be computed explicitly:
\al{
f^{(0)}_\alpha(t^*) &=  \frac{e^{-1/2 t^*}\left(1-t^* \left(3 t^*+4\right)\right) \Omega_\alpha}{256 \sqrt{2} \pi ^{5/2} \left(t^*\right)^{9/2}},\nl
f^{(1)}_{x/y}(t^*) &= \frac{e^{-1/2 t^*}\Omega _{y/x}}{512 \sqrt{2} \pi ^{5/2} \left(t^*\right)^{9/2} } \Big[1 +3 \left(t^*\right)^2\nl
&\qquad \qquad - \left(t^* \left(3 t^*+4\right)-t^*\right) \Omega _{x/y}^2\Big], \nl
f^{(1)}_z(t^*) &= \frac{e^{-1/2 t^*}\left(1-t^* \left(3 t^*+4\right)\right) \Omega _x \Omega _y \Omega _z}{512 \sqrt{2} \pi ^{5/2} \left(t^*\right)^{9/2} }.
\label{eq:Fscomponents}
}
In the following, $\Omega_\alpha\equiv \Omega_\alpha^{(\L0)}=L_{0,\alpha}/|\bm L_0|$ denotes the angular part of the components of $\bm L_0$ [see \Eref{eq:Omegadef}]. The functions $f^{(2)}_\alpha(t^*)$ are more cumbersome and are given in \Eref{eq:f2} in Appendix \ref{ap:FAppend}. Equation \eref{eq:FincStatExpand2} displays the power law dependences of the forces on $|\L0|$ and $\lambda$, with the limit $\lambda\gg|\L0|$ being implied throughout (recall the discussion of weak shear in Subsec.~\ref{sec:lengthscales}).

At late times after the quench, the components of the force decay algebraically in time:
\al{
f^{(0)}_\alpha(t^*\gg1)&=\frac{- 3 \Omega _\alpha}{256 \sqrt{2} \pi ^{5/2} \left(t^*\right)^{5/2}}, \nl
f^{(1)}_{x/y}(t^*\gg1)&=\frac{3 \left(1-\Omega _{x/y}^2\right) \Omega _{y/x}}{512 \sqrt{2} \pi ^{5/2} \left(t^*\right)^{5/2}},\nl
f^{(1)}_{z}(t^*\gg1)&=\frac{-3 \Omega _x \Omega _y \Omega _z}{512 \sqrt{2} \pi ^{5/2} \left(t^*\right)^{5/2}}.
\label{eq:flongtime}
}
From Eqs.~\eref{eq:FincStatExpand2} and \eref{eq:Fscomponents} we can construct the magnitude and the unit vector of each order contributing to the force. Introducing $\mathcal N(n) = {\alpha _1 \alpha _2 k_BT}/({|\bm L_0|^{7-2n}\lambda^{2n}})$, we find
\al{
\frac{|\bm F_{\textrm s}^{(0)}(t^*\gg1)|}{\mathcal N(0)} &= \frac{3}{2\left(8 \pi  t^*\right)^{5/2}},\nl
\frac{|\bm F_{\textrm s}^{(1)}(t^*\gg1)|}{\mathcal N(1)}&= \frac{3\;\sqrt{\Omega _x^2 \left(1-3 \Omega _y^2\right)+\Omega _y^2}}{4 \left(8 \pi  t^*\right)^{5/2}}, \nl
\frac{| \bm F_{\textrm s}^{(2)}(t^*\gg1)|}{\mathcal N(2)}&=\frac{3}{36 (8 \pi )^{5/2}\sqrt{t^*}}.
\label{eq:Flongtime}
}

While the zeroth- and first-order shear corrections decay as $\left(t^*\right)^{-5/2}$ at late times after the quench, the second-order correction decays more slowly as $\left(t^*\right)^{-1/2}$. This is a remnant of Eq.~\eqref{eq:CqxtModBLocal}, which shows that higher orders in shear are important at late times. Shear thus appears to dominate forces at late times. However, this regime is not accessible within the present approach. Indeed, \Eref{eq:CqxtModBLocal} (evaluated at $\bm x = \L0$) provides the condition for the crossover between the regimes of weak shear at short times and strong shear at late times, which must be satisfied for the expansion in \Eref{eq:FincStatExpand2} to hold. 

In Fig.~\ref{fig:f0F0} we show the time-dependent components of the force at zeroth order in shear. This limit corresponds to the result in Ref.~\cite{rohwer2017transient} for the post-quench force in a homogeneous system. Due to diffusing correlation fronts passing the inclusions~\cite{rohwer2017transient} (also see Fig.~\ref{fig:CQlocal}), the force changes sign at the reduced time
\al{
t^*_{\textrm{sgn}}=\frac{1}{3} \left(\sqrt{7}-2\right)\simeq0.215.
\label{eq:tsgn}
}
In the absence of shear, the force is parallel to the separation vector $\L0$. With shear, the force is modified. Figure~\ref{fig:F1s} shows the magnitude of the first shear correction. This correction depends sensitively on the orientation of $\L0$, and is maximal when $\L0$ lies in the $x$-$y$ plane, i.e., for $\vartheta=\pi/2$. Additionally, a larger separation along the $y$-axis implies a corresponding larger difference in shear velocity. The sign of this correction depends, inter alia, on the orientation of $\L0$, i.e., on $\vartheta$, $\varphi$, and $t^*$. Accordingly, the inclusions may experience an increase or decrease of the force due to shear, depending on their orientation relative to the shear plane.

The angle between $\bm L_0$ and $\bm F_{\textrm s}(\bm L_0,t)$ is determined by the unit separation and force vectors:
\al{
\cos\delta_{\textrm s}=\hat{\bm F}_{\textrm s}\cdot\hat{\bm L}_0\in[-1,1].
\label{eq:delta2s}
}
Using \Eref{eq:Fscomponents}, \Eref{eq:delta2s} is expanded in terms of powers of the shear rate $\dot\gamma = D/\lambda^2$:
\al{
\cos\delta_{\textrm s}&= \text{sgn}\left(t^* \left(3 t^*+4\right)-1\right)\Bigg[1 + \nl
&\qquad \frac{ \left(1-3 t^*\right)^2 \left(t^*\right)^2 \left(\Omega _x^2 \left(4 \Omega _y^2-1\right)-\Omega _y^2\right)}{8 \lamt ^4 \left(t^* \left(3 t^*+4\right)-1\right)^2}\Bigg]\nl
&\qquad \quad +\mathcal{O}(\lamt^{-6}).
\label{eq:delta2sresult}
}
The lowest order represents the change of sign as it occurs in the absence of shear [see \Eref{eq:tsgn}]. Due to symmetry, $\delta_{\textrm s}$ does not carry a first-order shear correction $\propto\lamt^{-2}$. The second-order shear correction in $\delta_{\textrm s}$ displays a singularity at $t^*=t^*_{\textrm{sgn}}$, associated with the divergence of the normalization of the unit vectors in \Eref{eq:delta2s}, so that the weak-shear expansion becomes invalid. 
The angle between the force and the vector connecting the inclusions depends on the orientation of $\L0$ as well as on time, because the post-quench correlations are distorted by shear and the magnitude of this distortion depends on the position of the inclusions in the shear field. Formally, one can compute the angle at long times from \Eref{eq:delta2s} by taking  $\lim_{t^*\to\infty}\hat{\bm F}_{\textrm s}\cdot\hat{\bm L}_0$; this renders~\footnote{
Equation \eref{eq:delta2nslongtime} does not follow from \Eref{eq:delta2sresult}, because the operations of expanding in terms of powers of shear and taking the limit $t^*\!\to\!\infty$ do not commute.
}
\al{
\lim_{t^*\!\to\!\infty}\cos\delta_{\textrm s}
=\Omega_x \textrm{sgn}(\Omega_y).
\label{eq:delta2nslongtime}
}
Thus the final angle of the force formed with $\L0$ appears to be independent of many details. However, the detailed study of this regime of late times requires investigative tools which go beyond those employed here. 

In summary, depending on the orientation of $\bm L_0$, the distortion of the post-quench correlations by shear can either increase or decrease the strength of the post-quench force between the inclusions. In addition, the angle between the separation vector of the inclusions  and the force gains a dependence on time (due to the evolution of sheared correlations) and on the orientation of the inclusions.

\subsubsection{Post-quench force between two inclusions advected by shear flow}
\label{sec:Fcomovsec}

\begin{figure*}[t]
\centering
\includegraphics[width=0.48\textwidth]{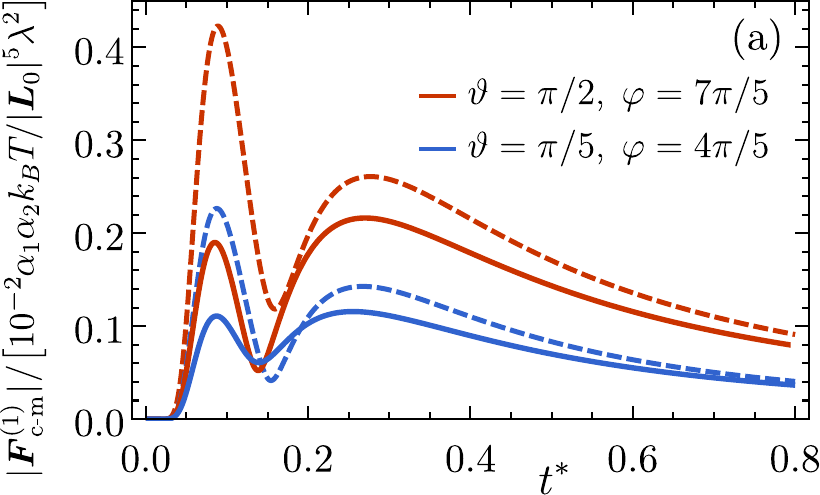}
\hspace{0.4cm}
\includegraphics[width=0.478\textwidth]{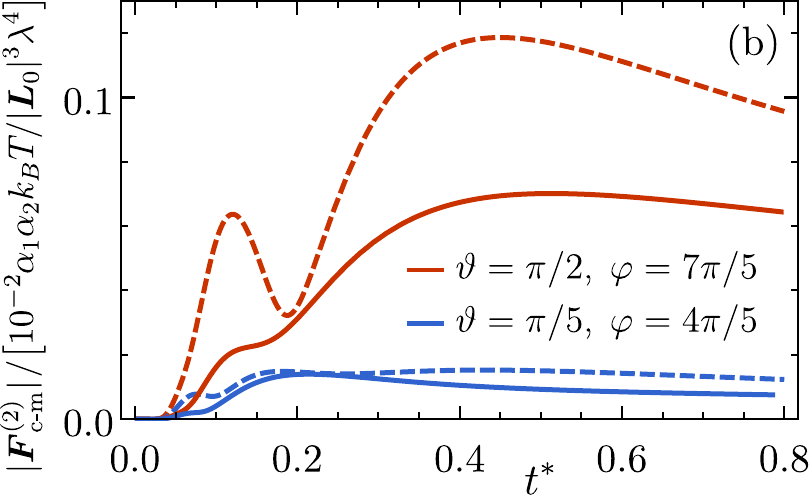}
%
\caption{Magnitudes of the first- (a) and second- (b) order shear correction of the post-quench force between two co-moving inclusions [connected by the vector $\Lt=\L0+\dot\gamma L_{0,y} t\bm e_x$] shown as functions of the rescaled time $t^* =D t/|\L0|^2$. Dashed lines correspond to the system in which the medium is unsheared [$|\bm F_{\textrm{c-m},0}|$ from Eqs.~\eref{eq:Fcm0def} and \eref{eq:Fcnoshearcomponents}], while solid curves represent the case of a sheared medium [$|\bm F_{\textrm{c-m},\dot\gamma}|$ from Eqs.~\eref{eq:Fcmsheardef} and \eref{eq:Fccomponents}]. The azimuthal $(\vartheta)$ and polar ($\varphi$) angles describe the orientation of the (initial) vector $\L0$ connecting the inclusions. 
}
\label{fig:F1F2cm}
\end{figure*}

In a typical experimental setup, the inclusion may be advected by the shear flow. In the following we shall study this scenario for two cases. In addition to the force $\bm F_{\textrm{c-m},\dot\gamma}(\bm L(t),t)\equiv\bm F_{\textrm{s}}(\bm L_0\to\bm L(t),t)$ between advected inclusions in the \textit{sheared fluid}, we shall also compute the force $\bm F_{\textrm{c-m},0}(\bm L(t),t)\equiv\bm F^{(0)}_{\textrm{s}}(\bm L_0\to\bm L(t),t)$ between two inclusions following the same advected trajectories, but in a system in which the correlated fluid is \textit{not sheared}. (In both cases the subscript ``c-m'' refers to co-moving inclusions.) This allows us to disentangle the effects of motion of the inclusions and those of shearing the post-quench correlations in the fluid. In both cases, the displacement vector is $\Lt=\L0+\dot\gamma L_{0,y} t\bm e_x$. As before, one has $\lamt=\lambda/|\L0|$ and $t^*=D t/|\L0|^2$. For the orders in shear, we obtain (see Fig.~\ref{fig:F1F2cm})
\al{
\bm F_{\textrm{c-m},0}^{(n)}&=
\frac{\alpha _1 \alpha _2 k_BT}{|\bm L_0|^{7-2n}\lambda^{2n}}\bm g^{(n)}(t^*)
\label{eq:Fcm0def}
}
and
\al{
\bm F^{(n)}_{\textrm{c-m},\dot\gamma}=
\frac{\alpha _1 \alpha _2 k_BT}{|\bm L_0|^{7-2n}\lambda^{2n}}  \bm h^{(n)}(t^*).
\label{eq:Fcmsheardef}
}
The components $\bm g^{(n)}_\alpha(t^*)$ and $\bm h^{(n)}_\alpha(t^*)$ can be obtained from $\bm f^{(n)}_\alpha(t^*)$ as given in Eqs.~\eref{eq:Fscomponents} and \eref{eq:f2}. For completeness, the explicit expressions for $n=0,1,2$ are provided in Eqs.~\eref{eq:Fcnoshearcomponents} - \eref{eq:h2} in Appendix \ref{ap:FAppend}. In those expressions, the quantity $\Omega_\alpha\equiv L_{0,\alpha}/|\bm L_0|$ and thus the angles are given with respect to the initial separation vector. At zeroth order in $\dot\gamma$, both forces in Eqs.~\eref{eq:Fcm0def} and \eref{eq:Fcmsheardef} naturally reduce to the force between stationary inclusions in an unsheared medium.

At the various orders of the shear expansion, the forces in Eqs.~\eref{eq:Fcm0def} and \eref{eq:Fcmsheardef} can also be decomposed into magnitudes and unit vectors. The results of this procedure are shown in Fig.~\ref{fig:F1F2cm} for the first- and second-order corrections to $|\bm F_{\textrm{c-m},0}|$ and $|\bm F_{\textrm{c-m},\dot\gamma}|$. These corrections clearly display a dependence on the initial orientation of $\Lt$ (i.e., the orientation of $\L0$ described in terms of $\vartheta$ and $\varphi$). The corrections are always maximal for $\vartheta=\pi/2$, i.e., if $\L0$  lies in the $x-y$ plane. We find that, at late times, $|\bm F_{\textrm{c-m},0}^{(1)}|$ and $|\bm F_{\textrm{c-m},\dot\gamma}^{(1)}|$ approach the same asymptotes [see \Eref{eq:Fcmlongtime}]. Indeed, the shear corrections of the forces acting on the co-moving inclusions relax more slowly at long times than the shear-free contribution. Explicitly we find [compare \Eref{eq:Flongtime}]
\al{
|\bm F_{\textrm{c-m},0}^{(0)}(t^*\gg1)|&= |\bm F_{\textrm s}^{(0)}(t^*\gg1)|\sim(t^*)^{-5/2},\nl
\frac{|\bm F_{\textrm{c-m},0}^{(1)}(t^*\gg1)|}{\mathcal N(1)}&= \frac{3  \sqrt{\left(3 \Omega _x^2+1\right) \Omega _y^2}}{256 \sqrt{2} \pi ^{5/2}\left({t^*}\right)^{3/2}}, \nl
\frac{| \bm F_{\textrm {c-m},0}^{(2)}(t^*\gg1)|}{\mathcal N(2)}&=\frac{9\sqrt{\left(5 \Omega _x^4-2 \Omega _x^2+1\right) \Omega _y^4}}{4 (8 \pi )^{5/2}\sqrt{t^*}}, \nl
|\bm F_{\textrm{c-m},\dot\gamma}^{(0)}(t^*\gg1)|&= |\bm F_{\textrm s}^{(0)}(t^*\gg1)|\sim(t^*)^{-5/2},\nl
|\bm F_{\textrm{c-m},\dot\gamma}^{(1)}(t^*\gg1)|&= |\bm F_{\textrm{c-m},0}^{(1)}(t^*\gg1)|\sim(t^*)^{-3/2}, \nl
\frac{| \bm F_{\textrm {c-m},\dot\gamma}^{(2)}(t^*\gg1)|}{\mathcal N(2)}&=\frac{1 }{1536 \sqrt{2} \pi ^{5/2}\sqrt{{t^*}}} \Big[54 \left(3 \Omega _x^2-1\right) \Omega _y^2 \nl
+ &729 \left(5 \Omega _x^4-2 \Omega _x^2+1\right) \Omega _y^4+1\Big]^{1/2} .
\label{eq:Fcmlongtime}
}


We conclude that at first order in shear, the force at late times is affected more strongly by the motion of the co-moving particles than by the shearing of the post-quench correlations. However, in the intermediate regimes, the two contributions differ with respect to their time-dependence. In turn, at second order, the correction to $|\bm F_{\textrm{c-m}}(t^*\gg1)|$ differs for the unsheared and the sheared system. This indicates that the combined effect of shearing and co-motion is visible at second order in shear at late times. Furthermore, shear corrections for the co-moving particles (for both the unsheared and the sheared system) relax more slowly at late times than those of the stationary inclusions [compare Eqs.~\eref{eq:Flongtime} and \eref{eq:Fcmlongtime}]. 
For both the unsheared and sheared system, successive orders in the shear corrections are longer-lived, decaying with ever increasing powers of $t^*$. As mentioned before, this indicates that the shear expansion is  invalid at late times [also see Eq.~\eqref{eq:CqxtModBLocal}].  Consequently we also expect the angle between the force and $\Lt$ to be determined by the shear corrections at late times, provided that $\lambda$ is large enough for the expansion to be valid in that regime.

Therefore we consider the angular dependence of these forces explicitly. The time-dependent unit vector connecting the two co-moving inclusions is 
$\hat{\bm L}(t)= \left(\frac{t^* \Omega _y}{\lamt^2}+\Omega _x,\Omega _y,\Omega _z\right)/\sqrt{1+2\Omega _x\Omega _y{t^* }/{\lamt^2}+ \left({t^* \Omega _y}/{\lamt^2}\right)^2}$.
Thus the angle $\delta_{\textrm{c-m},0/\dot\gamma}$ between the force and the inclusions is determined by
\al{
\cos\delta_{\textrm{c-m},0/\dot\gamma}=\hat{\bm F}_{\textrm{c-m},0/\dot\gamma}\cdot\hat{\bm L}(t).
\label{eq:deltacm}
}
For non-zero shear, the shear flow separates the inclusions along the $x$ axis [$\lim_{t^*\!\to\infty} \hat{\bm L}(t)=
\textrm{sgn}(\Omega _y)\bm e_x$]. [We note that the operations of switching off shear ($\lamt\!\to\!\infty$) and taking the late-time limit ($t^*\!\to\!\infty$) do not commute.] As in \Eref{eq:delta2sresult}, we expand the scalar product in \Eref{eq:deltacm} in orders of $\lamt^{-2}$, which renders
\al{
\cos \delta_{\textrm{c-m},0} = \text{sgn}\left(t^* \left(3 t^*+4\right)-1\right)
}
and
\al{
\cos \delta_{\textrm{c-m},\dot\gamma} &=\cos \delta_{\textrm{s}} + \mathcal{O}(\lamt^{-6}).
}
Therefore in the unsheared system, $\bm F_{\textrm{c-m},0}$ is always parallel to the vector $\Lt$ connecting the inclusions, which is a welcome cross check of our computations. In turn, at the expansion orders provided [compare \Eref{eq:delta2sresult}], the angle between $\Lt$ and $\bm F_{\textrm{c-m},\dot\gamma}$ is the same as the one between $\L0$ and $\bm F_{\textrm{s}}$.

\section{Conclusions and perspectives}
\label{sec:conc}

We have presented a systematic Gaussian study of spatial correlation functions as they occur after a quench in a sheared fluid. The quantity undergoing a quench could be either $k_B T/m$, i.e., the temperature and/or the compressibility of the fluid, or the correlation length $\xi$, or a combination of both. We have studied the sheared post-quench dynamics in the limit of small $\xi$, and as a function of the shear-induced length scale $\lambda=\sqrt{D/\dot\gamma}$. The presence (model B) or absence (model A) of the conservation of density fluctuations $\phi$ strongly influences correlations and forces. Our findings can be summarized as follows:

\begin{enumerate}[leftmargin=*]
 \item In a steady state, correlations under weak shear with dissipative dynamics decay as $e^{-1/\xi}$ for $\xi\!\to\!0$, as it is the case in equilibrium. In contrast, for conserved dynamics, the steady state correlation function displays long-ranged correlations which vanish algebraically for $\xi\!\to\!0$. Thus shear produces quantitative and qualitative corrections to correlation functions in systems with conserved dynamics. 
 \item Regarding shearing and quenching in systems with conserved dynamics, we observe long-ranged transient correlations, which are distorted by shear. Time-dependent correlation functions have been computed for various scenarios. (i) For vanishing $\xi$, we have obtained closed-form expressions, valid for all shear rates. Shear distorts the fronts of diffusively relaxing correlations (see Fig.~\ref{fig:CQlocal}), so that points can be more strongly or more weakly correlated than in an unsheared medium, depending on their displacement relative to the shear flow. (ii) Correlations between two points following an advected trajectory depend strongly on the initial displacement between the points (see Fig.~\ref{fig:Ccomov}). (iii) For non-zero values of $\xi$, the different contributions to the post-quench correlation function due to quenching $k_B T/m$ or $\xi$, as well as their dependence on weak shear, have been identified (see Fig.~\ref{fig:A(t)ModB}). At leading order, terms stemming from quenching $k_B T/m$ decay more slowly than those arising from a quench of $\xi$.
 \item We have extended the formalism of Ref.~\cite{deangopinathan2010pre} for computing post-quench fluctuation-induced forces, in order to include shear. This description applies to both the time-dependent and the steady state forces following a quench under shear, and can be used for a variety of geometries (e.g., parallel plates), thereby opening perspectives for numerous future research projects.  Here, the formalism was applied to the force between finite-sized inclusions (as sketched in Fig.~\ref{fig:IncSketch}), rendering a far-field force with properties resembling those of the aforementioned correlations. 
 \item In contrast to a homogeneous system, transient as well as steady state post-quench forces in a sheared medium are not parallel to the vector connecting the inclusions. Indeed, the forces depend strongly (both in magnitude and direction) on the (initial) relative orientation of the inclusions.
 \item In a steady state with weak shear, forces decay exponentially as $\xi\!\to\!0$ for model A, but algebraically for model B. In both cases, the orientation of the inclusions relative to the flow affects the magnitude as well as the direction of the force. The conservation law of the underlying dynamics therefore influences the observed non-equilibrium steady states; this differs strongly from equilibrium phenomena which are independent of the type of dynamics.
 \item Transient post-quench forces have been studied for the following cases: (i) static inclusions in a sheared medium, (ii) inclusions advected with the shear flow, and (iii) inclusions following advected trajectories in an unsheared medium. In the absence of shear, all cases reduce to the known result for a homogeneous system (see Fig.~\ref{fig:f0F0}). All forces are long-ranged, decaying algebraically with the (initial)  vector $\L0$ connecting the inclusions, as $\L0^{-7+2n}\lambda^{-2n}$ at the $n$th order in the shear rate $\dot\gamma$. Figures \ref{fig:F1s} and \ref{fig:F1F2cm} summarize the  angular and temporal dependence of the forces.
\end{enumerate}

We conclude that conservation laws play an important role in determining the character of non-equilibrium correlations. For conserved dynamics, quenches give rise to long-ranged effects, both in the transient and in the steady state regimes, even in the limit of small correlation lengths. If in addition the medium is sheared, strong spatial and orientational variations of fluctuation phenomena are observed. Based on this knowledge, correlations (and the associated fluctuation-induced forces) can be selectively enhanced or diminished.

The phenomena studied here are expected to have a large variety of experimental realizations, either for passive fluids under shear without a quench, or for active matter for which quenches can easily be introduced in addition. Indeed, non-equilibrium rheology is being explored experimentally and theoretically~\cite{BraderBallaufFuchs2010}. In particular, our findings are an important step toward harnessing the combination of fluctuation effects and shear in order to engineer interactions, e.g., between colloidal particles in correlated, quenched fluids. As far as physical realizations of quenches are concerned, suspensions of colloidal particles with tunable interactions~\cite{ballauff2006} are promising candidates.

Future studies may address the role of momentum conservation (corresponding to the so-called model H~\cite{hohenberg}). This would facilitate a connection to Refs.~\cite{dorfmankirkpatricksengers1994,Gompper2017shear,KirkpatrickSengers2018shear} which deal with fluctuation phenomena in hydrodynamic systems subject to shear. The above formalism can also be applied to forces in other geometries (e.g., thin films), so that other experimentally relevant setups (such as fluctuating wetting films) can be explored in the future. Extending the above formalism to time-dependent (e.g., oscillatory) shear would provide a further avenue for theoretical exploration and would potentially allow one to make contact with experiments~\cite{BraderBallaufFuchs2010}.

\begin{acknowledgments}
We thank M.~Gross for valuable discussions. 
During the preparation of this study, M.~Kr\"uger was supported by Deutsche Forschungsgemeinschaft (DFG) under the grants numbers KR 3844/2-1 and KR 3844/2-2.
\end{acknowledgments}

\appendix
\section{Extension of the formalism in Ref.~\cite{deangopinathan2010pre} to include shear}
\label{sec:deanappend}

In general, Gaussian Hamiltonians can be cast into the form
\al{
H = \frac 1 2 \int \textrm d^dx \,\textrm d^dx' \;\phi(\bx)\Delta(\bx,\bx',\bm L)\phi(\bx'),
}
so that, e.g., the Hamiltonian with inclusions [\Eref{Htot}] corresponds to the kernel
\al{
\Delta(\bx,\bx',\bm L) = \big[-\kappa\mathbf\nabla_{\bx}\cdot\mathbf\nabla_{\bx'} + m(\bx)\big]\delta(\bx-\bx'),
\label{eq_Deltatot}
}
where $m(\bx)$ is given in \Eref{eq_mx} and $\bm L$ is the separation vector of the inclusions. More generally, $\Delta$ may also incorporate boundary conditions for the surfaces of immersed objects \cite{deangopinathan2010pre}. The framework presented in Ref.~\cite{deangopinathan2010pre} can be used to compute forces between objects which can be cast in terms of $\Delta$. Then, as in \Eref{eq_Deltatot}, $\Delta(\bx,\bx',\bm L)$ gains a dependence on the separation vector $\bm L$ of the objects.
In thermal equilibrium, the force between the external objects can be computed from the partition function
\al{
\ave{F} = k_B T\nabla_{\bm L} \ln[Z(\Delta)],
\label{eq:FZ}
}
with $Z(\Delta) = \int \mathcal D \phi\; e^{-\frac \beta 2 \int\! \textrm d^dx \,\textrm d^d x'\phi(\bx)\Delta(\bx,\bx',\bm L)\phi(\bx')}$. 

Turning to dynamics, the Langevin equation with shear [see \Eref{eq:Langevin}] can be written as  
\al{
\partial_t \phi(\bx,t) &= - R(\bx,\bx')\frac{\delta H}{\delta \phi(\bx')} +S(\bx,\bx')\phi(\bx') + \eta(\bx,t) \nl
&=-R^{(\dot\gamma)}\Delta(\bx,\bx') \phi(\bx')+ \eta(\bx,t) 
\label{eq:langevinDG}
}
upon introducing the operator notation $AB(\bx,\bx')\equiv\int \textrm d^d x''\;A(\bx,\bx'')B(\bx'',\bx')$ (integration over repeated coordinates is implied). Here $R$ encodes dissipative or conserved dynamics for $\phi$ (i.e., model A or B), and can be mapped onto \Eref{eq:hatGamma} according to $R(\bx,\bx') = \hat\mu_{A/B}(\bx)\delta(\bx - \bx')$. Due to the fluctuation-dissipation theorem, $R$ also appears in the noise correlator:
\al{
\ave{\eta(\bx,t)\eta(\bx',t')} &= 2 k_B T \delta(t-t')R(\bx,\bx').
}
The operator $S(\bx,\bx') = \dot\gamma y\partial_x \delta(\bx - \bx ')$ represents the advection term for simple shear, and $R^{(\dot\gamma)} = R+S\Delta^{-1}$. 

For a given configuration of $\phi(\bx)$, the mean force can also be computed directly from the Hamiltonian:
\al{
\ave{F(t)} &= -\ave{\partial_L H} \nl
&= -\frac {k_B T} 2 \big[\nabla_{\bm L}\Delta(\bx,\bx',\bm L)\big]\Delta^{-1}(\bx,\bx',\bm L),
\label{eq:FdHdL}
}
due to the relation $C(\bx)=\ave{\phi(\bx)\phi(\bx')}=k_B T\Delta^{-1}(\bx,\bx',\bm L)$. Inspired by the analysis in Sec.~II A in Ref.~\cite{deangopinathan2010pre}, the equivalence of Eqs.~\eref{eq:FZ} and \eref{eq:FdHdL} can be exploited for instantaneous configurations of the field $\phi(\xt)$. First, we note that, for a quench from $T_I=0$ to $T$, the (temporal) Laplace transform of the correlation function $C(\xt)$ from \Eref{eq:Cxtgen} can be written as 
\al{
C (s) \!= \!\frac{k_B T}{s}\!\left[\Delta \!+ \!\frac{s (R^{(\dot\gamma)})^{-1}}{2} \right]^{-1}\! =\! \frac{k_B T}{s}(\Delta_s^{(\dot\gamma)})^{-1}.
\label{eq:Cs}
} 
This implies that, according to \Eref{eq:LF(s)}, the time-dependent (non-equilibrium) forces emerging after a quench can be computed from an \textit{effective equilibrium theory}.
For $\dot\gamma = 0$, our results reduce exactly to those of Ref.~\cite{deangopinathan2010pre}. However, the analysis holds also for $\dot\gamma\neq0$, because \Eref{eq:langevinDG} is still a \textit{linear} Langevin equation and $S$ has a local kernel $\propto\delta(\bx - \bx ')$ for simple shear, i.e., $\dot\gamma=\,$const. From $\Delta_s^{(\dot\gamma)}$ [see below \Eref{eq:LF(s)}] one infers that the source of LRCs can be either the inherent correlations manifest in the Hamiltonian (via $\Delta$), or the presence of a conservation law (via $R$).

As stated, the above arguments also apply to the force acting between two inclusions separated by a vector $\bm L$. In thermal equilibrium, inclusions induce an additional contribution 
$\Delta \mathcal{F} = k_B T \ln \frac{Z_H}{Z_{H-H_{\textrm{inc}}}}$
to the free energy of the system, with the total and inclusion Hamiltonians $H$ [Eqs.~\eref{Htot}] and $H_{\textrm{inc}}$ [\Eref{Hinc}], respectively, and where
$Z_H = \int \mathcal D \phi\; e^{-\beta H[\phi]}$. 
For $|\bm L|\gg V_i^{1/d}$, an effective potential between the inclusions can be constructed via a cumulant expansion, which yields, after some Wick contractions,
\al{
\mathcal V(\bm L,t) = \frac{k_B T c_1 c_2 V_1 V_2}{2}\ave{\phi(\bm 0,t) \phi(\bm L,t)}^2.
}
This is in line with the arguments employed for computing equilibrium thermal Casimir forces between quadratically coupled inclusions in a near-critical fluid (see, e.g., Refs.~\cite{eisenriegler1995sphericalcrit,eisenriegler1995sphericalcritPRL,hankeeisenrieglerdietrich1998Spherical}). However, because $H_{\textrm{inc}}$ is Gaussian, too, the above Laplace transform formalism can be applied in order to compute the (time-dependent) non-equilibrium potential after a quench. For $L\gg V_i^{1/d}$, \Eref{eq:VLt} is exactly recovered.

\onecolumngrid

\section{Shear corrections to forces between inclusions}
\label{ap:FAppend}
Below we provide the contributions to the shear rate expansion of the forces between two inclusions, as discussed in Sec.~\ref{sec:incsec}. For stationary inclusions immersed in a sheared fluid with post-quench correlations, $\bm F_{\textrm s}^{(2)}$ in \Eref{eq:FincStatExpand2} has the following vector components:
\al{
f^{(2)}_x(t^*) 
&=\frac{e^{-\frac{1}{2 t^*}} \Omega _x \left(3 \Omega _y^2 \left(\left(1-t^* \left(3 t^*+4\right)\right) \Omega _x^2+t^* \left(t^* \left(t^*+8\right)-3\right)\right)-t^* \left(\left(t^* \left(t^*+2\right)-1\right) \Omega _x^2+4 \left(\left(t^*\right)^3+t^*\right)\right)\right)}{6144 \sqrt{2} \pi ^{5/2} \left(t^*\right)^{9/2}},\nl
f^{(2)}_y(t^*)
&= \frac{e^{-\frac{1}{2 t^*}} \Omega _y \left(t^* \left(\left(t^*\right)^2+2 t^*-1\right) \left(3 \Omega _y^2-4 t^*\right)
-\Omega _x^2 \left(3 \left(3 \left(t^*\right)^2+4 t^*-1\right) \Omega _y^2+t^* \left(\left(t^*\right)^2-16 t^*+5\right)\right)
\right)}{6144 \sqrt{2} \pi ^{5/2} \left(t^*\right)^{9/2}},\nl
f^{(2)}_z(t^*) 
&= \frac{e^{-\frac{1}{2 t^*}} }{6144 \sqrt{2} \pi ^{5/2} \left(t^*\right)^{9/2}} \Omega_z \Big(
3 \Omega _y^2 \left(\left(1-t^* \left(3 t^*+4\right)\right) \Omega _x^2+t^* \left(t^* \left(t^*+2\right)-1\right)\right) \nl
&\qquad\qquad\qquad\qquad\qquad\qquad-t^* \left(\left(t^* \left(t^*+2\right)-1\right) \Omega _x^2+2 t^* \left(2 \left(t^*\right)^2+t^*+1\right)\right)
\Big).
\label{eq:f2}
}
For inclusions following the trajectory of a shear flow, embedded in an unsheared fluid with post-quench forces, $\bm F_{\textrm{c-m},0}^{(n)}$ in \Eref{eq:Fcm0def} has the vector components ($\alpha = x,y,z$)
\al{
g^{(0)}_\alpha(t^*) &= f^{(0)}_\alpha(t^*), \nl
g^{(1)}_{x}(t^*) &= \frac{e^{-\frac{1}{2 t^*}} \Omega _y }{256 \sqrt{2} \pi ^{5/2} \left(t^*\right)^{9/2}} \Big(
\left(3 t^* \left(2 \left(t^*\right)^2+t^*+2\right)-1\right) \Omega _x^2-t^* \left(t^* \left(3 t^*+4\right)-1\right) \left(\Omega _y^2+\Omega _z^2\right)
\Big)\nl
g^{(1)}_y(t^*) 
&= \frac{e^{-\frac{1}{2 t^*}} \left(t^* \left(t^* \left(9 t^*+7\right)+5\right)-1\right) \Omega _x \Omega _y^2}{256 \sqrt{2} \pi ^{5/2} \left(t^*\right)^{9/2}},\nl
g^{(1)}_z(t^*) 
&= \frac{e^{-\frac{1}{2 t^*}} \left(t^* \left(t^* \left(9 t^*+7\right)+5\right)-1\right) \Omega _x \Omega _y \Omega _z}{256 \sqrt{2} \pi ^{5/2} \left(t^*\right)^{9/2}}
\label{eq:Fcnoshearcomponents}
}
at zeroth and first order in shear, respectively, and
\al{
g^{(2)}_x(t^*) 
&=\frac{e^{-\frac{1}{2 t^*}} \Omega _x \Omega _y^2 \left(3 t^* \left(t^* \left(t^* \left(9 t^*+7\right)+5\right)-1\right)-\left(3 t^* \left(t^* \left(5 t^* \left(3 t^*+2\right)+4\right)+2\right)-1\right) \Omega _x^2\right)}{512 \sqrt{2} \pi ^{5/2} \left(t^*\right)^{9/2}}, \nl
g^{(2)}_y(t^*) 
&= \frac{e^{-\frac{1}{2 t^*}} \Omega _y^3 \left(t^* \left(t^* \left(t^* \left(9 t^*+7\right)+5\right)-1\right)-\left(3 t^* \left(t^* \left(5 t^* \left(3 t^*+2\right)+4\right)+2\right)-1\right) \Omega _x^2\right)}{512 \sqrt{2} \pi ^{5/2} \left(t^*\right)^{9/2}}, \nl
g^{(2)}_z(t^*) 
&= \frac{e^{-\frac{1}{2 t^*}} \Omega _y^2 \Omega _z \left(t^* \left(t^* \left(t^* \left(9 t^*+7\right)+5\right)-1\right)-\left(3 t^* \left(t^* \left(5 t^* \left(3 t^*+2\right)+4\right)+2\right)-1\right) \Omega _x^2\right)}{512 \sqrt{2} \pi ^{5/2} \left(t^*\right)^{9/2}}
\label{eq:g2}
}
at second order in shear. Lastly, for comoving inclusions embedded in the system with correlations subject to shear, $\bm F_{\textrm{c-m},\dot\gamma}^{(n)}$ in \Eref{eq:Fcmsheardef} has the vector components
\al{
h^{(0)}_\alpha(t^*) &= f^{(0)}_\alpha(t^*), \nl
h^{(1)}_{x}(t^*) 
&= \frac{e^{-\frac{1}{2 t^*}} \Omega _y \left(\left(t^* \left(t^* \left(18 t^*+11\right)+6\right)-1\right) \Omega _x^2-t^* \left(t^*+1\right) \left(6 t^*-1\right)\right)}{512 \sqrt{2} \pi ^{5/2} \left(t^*\right)^{9/2}},\nl
h^{(1)}_y(t^*) 
&=- \frac{e^{-\frac{1}{2 t^*}} \Omega _x \left(t^* \left(-\left(t^* \left(18 t^*+11\right)+6\right) \Omega _y^2-3 t^*+1\right)+\Omega _y^2\right)}{512 \sqrt{2} \pi ^{5/2} \left(t^*\right)^{9/2}},\nl
h^{(1)}_z(t^*) 
&= \frac{e^{-\frac{1}{2 t^*}} \left(t^* \left(t^* \left(18 t^*+11\right)+6\right)-1\right) \Omega _x \Omega _y \Omega _z}{512 \sqrt{2} \pi ^{5/2} \left(t^*\right)^{9/2}}
\label{eq:Fccomponents}
}
at zeroth and first order in shear, respectively, and
\al{
h^{(2)}_x(t^*) 
&= \frac{-e^{-\frac{1}{2 t^*}} \Omega _x }{6144 \sqrt{2} \pi ^{5/2} \left(t^*\right)^{9/2}}\Bigg\{ 
3 \Omega _y^2 \left[\left(t^* \left(t^* \left(12 t^* \left(15 t^*+7\right)+23\right)+8\right)-1\right) \Omega _x^2-t^* \left(t^* \left(t^* \left(108 t^*+49\right)+20\right)-3\right)\right] \nl
&\qquad \qquad \qquad \qquad \qquad \qquad + t^* \left(\left(t^* \left(t^*+2\right)-1\right) \Omega _x^2+4 \left(\left(t^*\right)^3+t^*\right)\right) 
\Bigg\},\nl
h^{(2)}_y(t^*) 
&=\frac{-e^{-\frac{1}{2 t^*}} \Omega _y }{6144 \sqrt{2} \pi ^{5/2} \left(t^*\right)^{9/2}}\Bigg\{ 
3 \Omega _y^2 \left[\left(t^* \left(t^* \left(12 t^* \left(15 t^*+7\right)+23\right)+8\right)-1\right) \Omega _x^2-t^* \left(t^* \left(t^* \left(36 t^*+17\right)+6\right)-1\right)\right]\nl
&\qquad \qquad \qquad \qquad \qquad \qquad +t^* \left(\left(t^* \left(37 t^*+32\right)-7\right) \Omega _x^2+4 t^* \left(\left(t^*-7\right) t^*+2\right)\right)
\Bigg\},\nl
h^{(2)}_z(t^*) 
&= \frac{-e^{-\frac{1}{2 t^*}} \Omega _z }{6144 \sqrt{2} \pi ^{5/2} \left(t^*\right)^{9/2}}\Bigg\{ 
3 \Omega _y^2 \left[\left(t^* \left(t^* \left(12 t^* \left(15 t^*+7\right)+23\right)+8\right)-1\right) \Omega _x^2-t^* \left(t^* \left(t^* \left(36 t^*+17\right)+6\right)-1\right)\right]
\nl
&\qquad \qquad \qquad \qquad \qquad \qquad +t^* \left(\left(t^* \left(t^*+2\right)-1\right) \Omega _x^2+2 t^* \left(2 \left(t^*\right)^2+t^*+1\right)\right)
\Bigg\}
\label{eq:h2}
}
at second order in shear.

\twocolumngrid

\pagebreak

\bibliographystyle{apsrev4-1}
\bibliography{references}

\end{document}